\documentclass{aa}
\usepackage{graphicx}
\usepackage{natbib}
 \usepackage{aalongtable}
\usepackage{txfonts}
\begin{document}
   \title{Abundances in bulge stars from high-resolution, near-IR spectra\\ I. The CNO elements observed during the {\it science verification} of  CRIRES at VLT
    \thanks{Based on observations collected at the European Southern Observatory, Chile 
(ESO Programme 60.A-9058A)} 
}

   \author{Nils Ryde
          \inst{1,2}
         \and
         Bengt Edvardsson
          \inst{2}
         \and
         Bengt Gustafsson
          \inst{2}
         \and
          Kjell Eriksson
          \inst{2}
         \and
       Hans Ulrich K\"aufl\inst{3}
         \and
       Ralf Siebenmorgen\inst{3}
         \and
       Alain Smette\inst{4}
          }

   \offprints{N. Ryde}

   \institute{Lund Observatory, Box 43, SE-221 00 Lund, Sweden
   \and
   Department of Astronomy and Space Physics, Uppsala University, Box 515, SE-751 20 Uppsala,
         Sweden
         \and
ESO, Karl-Schwarzschild-Str. 2, 85748 Garching, Germany
\and
ESO, Alonso de Cordova 3107, Vitacura, Casilla 19001, Santiago 19, Chile\\
         \email{ryde@astro.lu.se}
             }

   \date{Received ; accepted }


  \abstract
   {The formation and evolution of the Milky Way bulge is not yet well understood and its classification is ambiguous. Constraints can, however,  be obtained by studying the abundances of key elements in bulge stars.}
   {The aim of this study  is to determine the chemical evolution of C, N, O, and a few other elements in stars in the Galactic bulge, and to  discuss  the sensitivities of the derived abundances from molecular lines.}
   { High-resolution, near-infrared spectra in the H band were recorded using the
   CRIRES spectrometer on the {\it Very Large Telescope}. Due to the high and variable visual extinction in the line-of-sight towards the bulge, an analysis in the near-IR is preferred. The C, N, and O abundances can all be determined  simultaneously from the numerous molecular lines in the wavelength range observed.}
   {The three giant stars in Baade's window presented here are the first bulge stars observed with CRIRES during its {\it science verification observations}. We have especially determined the C, N, and O abundances, with uncertainties of less than 0.20 dex,  from CO, CN, and OH lines.  Since the systematic uncertainties in the derived C, N, and O abundances due to uncertainties in the stellar fundamental parameters, notably $T_\mathrm{eff}$,  are significant, a detailed discussion of the sensitivities of the derived abundances is included.  We find  good agreement between near-IR and optically determined O, Ti, Fe, and Si abundances.  Two of our stars show a solar [C+N/Fe], suggesting that these giants have experienced the first dredge-up and that the oxygen abundance should reflect the original abundance of the giants. The two giants fit into the picture, in which  there is no significant difference between the oxygen abundance in bulge and thick-disk stars. Our determination of the sulphur abundances is the first for bulge stars. The high [S/Fe] values for all the stars indicate a high star-formation rate in an early phase of the bulge evolution. 
   }
   {}


   \maketitle
%

\section{Introduction}

The origin and chemical properties of the Galactic bulge are poorly understood, see, for instance, the reviews by \citet{wyse} and Kormendy \& Kennicutt (2004)\nocite{kormendy}. These properties
are critical for our understanding of the formation and evolution of
the Milky Way, but also of galaxies in general \citep{renzini}.

  \begin{figure*}
   \centering
 \includegraphics[ width=\textwidth]{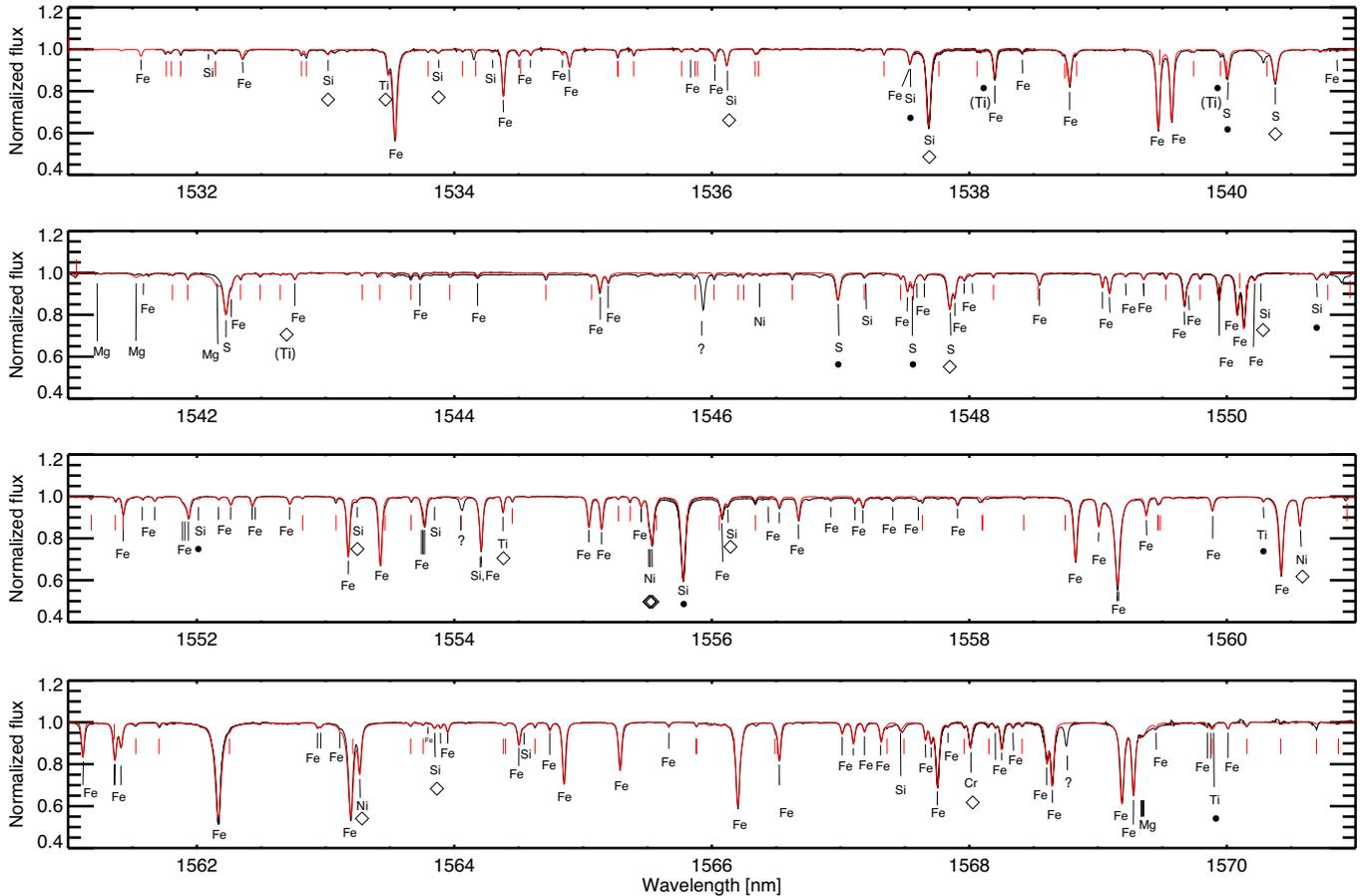}
      \caption{Observed Sun spectrum  \citep{solar_IR_atlas}  in the H band is shown with a full, black line. Our best synthetic spectrum is shown in red. All synthetic lines which are deeper than 0.97 of the continuum are marked.  Metal lines are shown below the spectrum. The metal lines are fitted to the solar atlas by changing their line strengths given the abundances compiled by \citet{solar2}. The lines used for an abundance analysis in the stellar spectra are shown with diamonds (for good unblended lines) and dots (for blended, slightly less sensitive lines). A few of the good lines are too strong and therefore very sensitive to the microturbulence and less useful for determining the corresponding abundance. A few Ti lines which are seen in our stellar spectra are not visible in the solar spectrum.  CN lines are indicated below the spectrum by red lines.}
         \label{sun}
   \end{figure*}

The classification of the Galactic bulge is also ambiguous: Its structure and shape, with a peanut-formed profile and 
a small bar, dynamic and transient in nature, show signatures
of the secularly evolved `pseudo-bulges'. Binney (ref IAU254, in press) defines it as a clear case of a pseudo-bulge. These are typical for late-type
galaxies, containing young stars, and believed to form slowly out of
disk gas and stars \citep{kormendy}. 

However, the stars in the bulge seem to show
ages and enhancements of $\alpha$ elements\footnote{The $\alpha$ elements include O, Mg, Si, S, Ca, and Ti.} relative to iron sooner characteristic of
`classical bulges', suggesting that the
star-formation period was early and very short \citep{lecureur,fulbright:07}. The classical bulges are typical for Sa and Sb galaxies,
they are similar to elliptical galaxies, and are interpreted as formed through hierarchical clustering and merging events. Furthermore, they probably had most of their star formation
long ago and therefore contain mostly old stars.  Indeed, colour-magnitude diagrams indicate that most
of the Galactic bulge stars formed more than $10$ Gyr ago \citep{ortolani} and \citet{zoccali:2003} 
find no trace of a younger population in the bulge. 
Hence, the bulk of the bulge's stellar population is thought not to have
formed by slow secular evolution. For instance, \cite{minniti:08} conclude in their review on the Galactic bulge, that a secular evolution of the disk forming a bulge can 
be excluded.\footnote{Although there is a considerable agreement that there exists a dominant, old, metal-rich
population in the bulge,  there is some evidence of ongoing star formation and a younger population within of the order of 100 pc of the geometrical centre, see e.g. \citet{figer}  and \citet{barbuy}.}

Hence, the formation of the Milky Way bulge is not well understood and its classification is still inconclusive. Since the different formation scenarios can be
constrained by abundance surveys, more of them need to be performed.  The two types of bulges give different dynamic and chemical signatures. From the $\alpha$-element compositions relative to iron as a function of the metallicities, [Fe/H], of the stars, one can infer star-formation rates (SFR) and initial-mass functions (IMF) \citep[see, for example,][]{carnegie}. A shallower IMF will increase the number of $\alpha$-element producing stars thus leading to higher [$\alpha$/Fe] values. A faster enrichment due to a high star-formation rate will keep the over-abundance of the $\alpha$ elements relative to iron at a high value also at higher metallicity. Different populations may show different behaviours.

To fully clarify the situation, a determination of the degree of mixing-in of other populations than the dominant one needs to be done. A search for other populations should be performed in different parts of the bulge, especially in the centre and along the galactic plane, by abundance surveys of bulge stars. Until now, the bulge
has remained fairly unexplored, mainly because of the high and variable optical obscuration due to dust
in the line-of-sight toward the Galactic centre.  Other difficulties that affect investigations of bulge stars are the crowding of stars and the confusion of foreground stars. 
However, the infrared, with lower
extinction ($\mathrm{A_{K} \sim 0.1 \times A_V}$; Cardelli et al. (1989))\nocite{cardelli} and predominantly molecular rather than atomic abundance
indicators, is a preferred wavelength region to study abundances in
bulge stars \citep{ryde_munchen_review}. This will allow the investigation of the entire bulge and not only regions with low optical obscuration. Discussions of abundances based on high-resolution spectroscopy in the IR are becoming more common due to spectrometers such as the Phoenix  \citep{phoenix,phoenix:2003} and CRIRES spectrometers \citep{crires1,crires2}. In this paper, which is the first in a series, we have studied abundances in the first three bulge giants based on near-IR  spectra observed with the CRIRES spectrometer, during its science Verification observations.  We have in particular studied the abundances of  the C, N, and O elements, as derived from molecular lines, and discuss their uncertainties.

  \begin{figure*}
   \centering
 \includegraphics[ width=\textwidth]{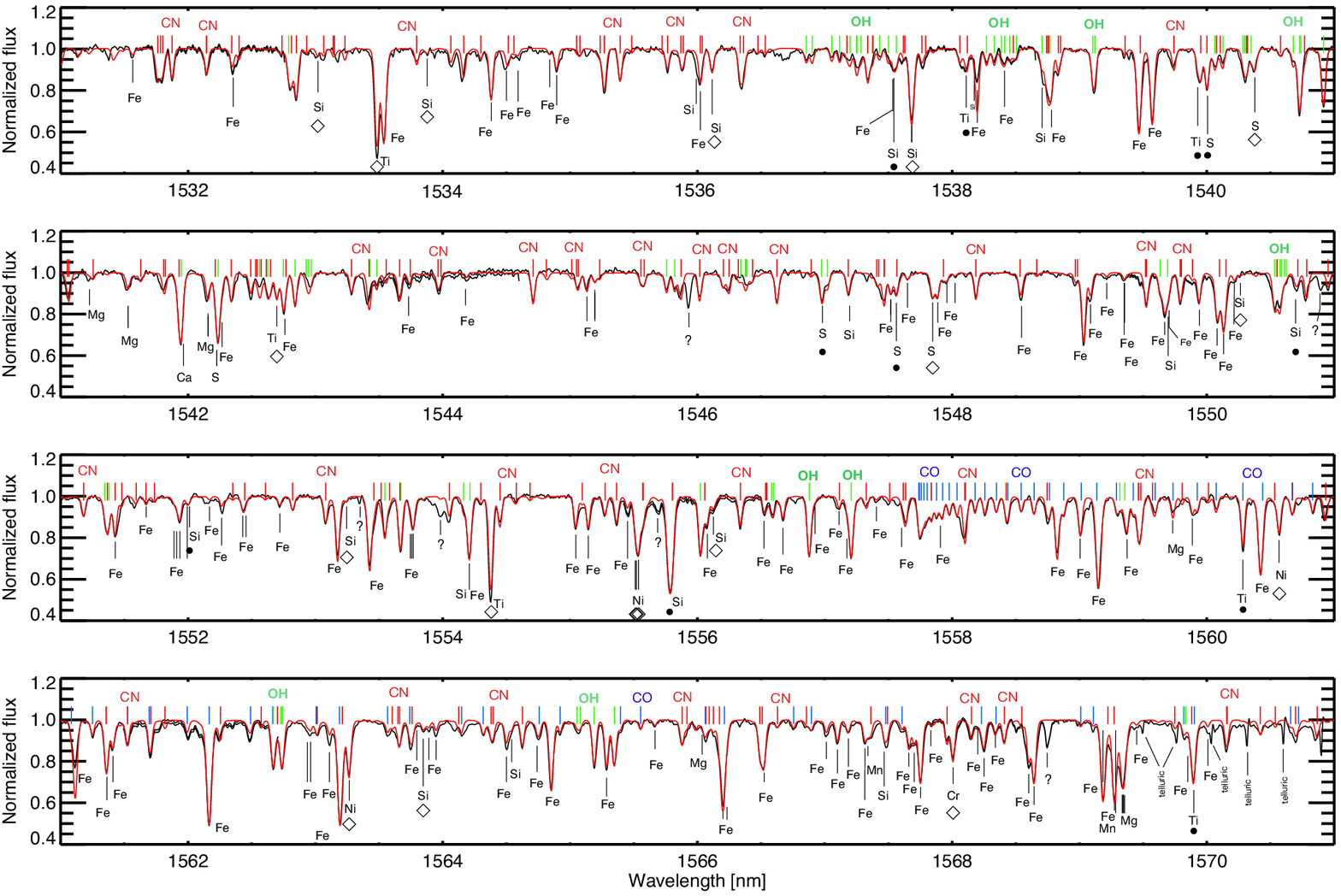}
      \caption{Observed Arcturus spectrum  \citep{arcturusatlas_II}  is shown with a full, black line. Our best synthetic spectrum is shown in red. All synthetic lines which are deeper than 0.97 of the continuum are marked.  Metal lines are indicated below the spectrum. A lot of them are blended with molecular lines and some are not visible as such, but these are marked anyway since they do contribute to the spectral feature. The $gf$values of metal lines visible in the solar spectrum are determined from
Fig. \ref{sun}. The other lines are fitted in the Arcturus atlas by changing their line strengths given the abundances deduced by \citet{fulbright:07}. 
Molecular lines are indicated above the spectrum (red for CN, blue for CO and green for OH). Some well-suited molecular lines for an abundance determination (i.e. of suitable strength and unblended) are indicated with CN, CO, and OH respectively. As can be seen the weak and strong OH lines can not be fitted with the same abundance. The stronger OH($v=2-0, 3-1$) lines indicate an abundance of $A_O=8.79$ and the weaker OH($v=4-2$) ones require an oxygen abundance  0.1 dex lower  to be fitted.  This difference is not understood but will be investigated later. }
         \label{aboo}
   \end{figure*}

\section{Abundance determinations: going to the near-IR}

In the wavelength range recorded by CRIRES, we have identified lines from which abundances can be derived for 
C, N, O,  the
$\alpha$ elements  Si, S, Ti, and the iron peak elements Cr, Fe, and Ni.  In particular, the abundances of the important C, N, and O elements can be better determined from a wealth of molecular lines in the 
near-IR, than in the optical. Apart from the fact that the extinction towards the bulge is lower at infrared wavelengths, there are other advantages of going to the near-IR.

\subsection{Advantages and drawbacks of going to the near-IR}

\label{sectabun}

\begin{enumerate}

\item Red stars stand out the most in the near-IR not only because they are brightest there.
Admittedly, the dust extinction decreases monotonically with wavelength, which favours longer wavelengths, but
interstellar dust radiates strongly in the mid- and far-IR, and therefore the near-IR should be preferred. Furthermore, in the thermal infrared
(i.e. beyond
approximately 2.3 $\mu$m) the telluric sky and the telescope shine due to their intrinsic temperatures, making observations increasingly difficult, the longer the wavelength. Thus, this leaves us with the J, H, and K bands for an optimal spectroscopic study.
 \item The near-IR is also preferred to the visual wavelength region for analysis of abundances due to
the fact that the absorption spectra are less crowded with lines, that fewer lines are blended, and that it is
easier to find portions of the spectrum which can be used to define a continuum. These are important facts which  reduce the uncertainties in the derived abundances. 
 \item For the rotation-vibration bands in the IR that occur within the electronic gound state, the assumption of \emph{local thermodynamic equilibrium (LTE)}
in the analysis of the molecules is probably valid \citep{hinkle_lambert:75}, 
which simplifies the correct analysis dramatically.
\item  Moreover, in the Rayleigh-Jeans regime, the continuum
intensity is less sensitive to temperature variations. This means that
the effects of, for example, effective-temperature uncertainties or
surface inhomogeneities on line strengths should be smaller in the
IR. \footnote {Note, however, that this effect also weakens the lines correspondingly.} 
\item  Not least,  it is important to realise that in general the
determination of the oxygen abundance can only be carried out with
confidence if the carbon abundance is well known, since the CO
molecule holds most of the C and much of O in cool star atmospheres. Only the IR offers, even within a small wavelength range,
all indicators necessary to accurately determine the C-N-O molecular
and atomic equilibrium in the atmospheres of cool stars, through the
simultaneous observation of many clean CO, CN and OH lines.
\end{enumerate}

A general drawback of a spectral analysis in the near-IR is that there are much fewer atomic and
ionic lines. The ones that exist often originate
from highly excited levels in metal atoms, which also complicates
an interpretation. Furthermore, many lines are not properly
identified and/or lack known oscillator strengths, which are
needed in an abundance analysis. 
Furthermore, even though great advances have been made in the technology for recording near-infrared light, existing
spectro\-meters are still much less effective than optical ones, one
of the main reasons being the lack of cross-dispersion.
Finally, determining the stellar parameters based only on near-IR spectra is difficult. 

Nevertheless, the near-IR is the optimal spectroscopic region to work in, in order
to get a handle on the Milky Way bulge through abundance analysis.
Recently,  a few studies of elemental abundances of bulge stars using near-IR spectra at high resolution ($R\sim50,000$) have been performed with the Phoenix spectrometer, see for instance  Mel\'endez et al. (2003), Cunha \& Smith (2006), Cunha et al. (2007), and Mel\'endez et al. (2008). 
\nocite{melendez:2008} \nocite{cunha:2007:apj} \nocite{cunha:2006}\nocite{melendez_gemini} 
Interesting investigations of bulge stars have also been performed with NIRSPEC at the Keck II telescope at 1.5-1.8~$\mu$m, see for instance \citet{origlia_GC4,origlia:2003,origlia_GC3,origlia_M,rich:2007}. However, in these studies, the spectral resolution is lower, $R\sim25,000$, which does not allow stellar lines to be fully resolved. 

\section{Observations}

In the project `CNO abundances in bulge giants',  
we were granted 3 hours of science verification observations with CRIRES. CRIRES is a cryogenic echelle
spectrograph designed for  high spectral resolution,
near-infrared observations.  This paper is based on the spectra obtained during those observations, which were performed on 12 August 2006. Adaptive optics (MACAO - Multi-Applications Curvature Adaptive Optics) was used, which enhances the signal-to-noise ratio and the spatial resolution. The stars observed are
Arp\,4329, Arp\,4203, and Arp\,1322\footnote{The order of the stars given here and throughout the paper is based on the strengths of the CO and OH bands, which can be seen in Figure \ref{obs_model} }  in Baade's Window. These stars were chosen among the bulge giants observed optically by \citet{fulbright:06}. The $H$ magnitudes and the total exposure times, $t_\mathrm{exp}$, for each programme star are given in Table \ref{obslog}.

The slit width was 0.4\arcsec\  yielding a spectral resolution of  $R = \lambda/\Delta \lambda = 50,000$ with four pixels per spectral resolution element. 
The wavelength range expected 
was $1532.6-1570.5$~nm (in order 36) over 
the detector arrays, consisting of a mosaic of four Aladdin III InSb arrays in the focal plane.
However, both  the first and fourth detector arrays had problems.\footnote{During early science verification, detectors 1 and 4 suffered some
vignetting. Moreover, a quasi `odd-even'-effect, characteristic of this type
of array, could not yet be mitigated by non-linear flatfielding. In the
future, within the limits of the width of the Echelle order, also these
detectors can be employed for precison abundance studies and to constrain the
stellar continuum.} Thus, only the spectra from the second and third arrays have been used for the analysis. Data for the first and fourth ones were used to confirm abundances when possible. This is, nevertheless, an improvement in wavelength coverage compared with the Phoenix detector, which corresponds to circa one of CRIRES's detector arrays ($\Delta \lambda = 0.5\%$). The approximate signal-to-noise ratios (SNR)  per pixel-element in the dispersion direction (the spectrum collapsed in the spatial direction) at the continuum at 1554.8 nm of the observed spectra (in the third detector array), are also given in Table \ref{obslog}. The SNR per wavelength resolution element is a factor of 2 larger.
The SNR varies by a factor of two between the detector arrays. The SNR of the second array varies approximately from 70\% to 
100\% linearly with wavelength, whereas  the third detector array,  which is the best one, provides a relatively constant SNR. 






The observed data were processed with standard routines in the reduction package
{\tt IRAF} \citep{IRAF}\footnote{IRAF is distributed by the National Optical Astronomy Observatory, which is operated by the Association of Universities for Research in Astronomy (AURA) under cooperative agreement with the National Science Foundation.}, in order  to retrieve one-dimensional, continuum normalised, and
wavelength calibrated stellar spectra. The four arrays were reduced separately, resulting in four spectra, one for each array.
In order to remove the detector dark current and the background emission from the thermal sky and telescope from the stellar frames, the telescope
was nodded along the slit between the source and a region of the sky (in an `AB' configuration with a nod throw of $10\arcsec$). The two equally long exposures
were subtracted from each other, yielding two sky-subtracted  exposures of the star. The reduced spectra from these exposures, were 
combined in {\tt IRAF} with task {\tt scombine}. In the wavelength region observed, the sky is relatively free from 
telluric lines, so no division by a standard star was made. 
The wavelength calibration was performed using the stellar lines themselves.
The local continua of the spectra were fitted and
normalised by a 4th-order Legendre function with the {\tt IRAF} task {\tt continuum}. 

Irrespective of the width of the entrance slit - if one opens it to more than the width  (FWHM) 
corresponding to the stellar image -  the factual entrance slit becomes the star itself. So for
our case, the effective spectral resolution was higher then $R=50,000$, but not
well constrained. Thus for the sake of
consistency all spectra are degraded to the worst case, that is a
spectral resolution of $R=50,000$.

\begin{table}
  \caption{Log of our observations. The stars are taken from \citet{arp}. }
  \label{obslog}
  \begin{tabular}{c c c c c c}
  \hline
  \noalign{\smallskip}
    Star & RA & dec &  $H$ & $t_\mathrm{exp}$ & SNR \\
  \noalign{\smallskip}
  \hline
  \noalign{\smallskip}
 \textrm {Arp\,4203} &  18 \,03\, 23.6 &  -30 \,01\, 59  & 9.2  & 200 s & 80\\
  \noalign{\smallskip}
 \textrm {Arp\,4329} & 18 \,03\, 28.4  & -29 \,58 \,42 & 11.1 & 1800 s & 95\\
  \noalign{\smallskip}
 \textrm {Arp\,1322} &  18 \,03 \,49.4 & -30 \,01 \,54   & 10.3 & 600 s  & 110 \\
 \noalign{\smallskip}
  \hline
  \end{tabular}
\end{table}

\begin{table*}
  \caption{Abundances  for Arcturus and the three bulge giants. }
  \label{abundances1}
  \begin{tabular}{l l c c c c c c c c c c}
  \hline
  \noalign{\smallskip}
    Star &  Ref. & $\mathrm { \log\varepsilon(C) }$$^{\mathrm{a}}$   & $\mathrm { \log\varepsilon(N) }$ & $\mathrm { \log\varepsilon(O) }$  & $\mathrm { \log\varepsilon(Ti) }$ & $\mathrm {  \log\varepsilon(Si)}$ & $\mathrm {  \log\varepsilon(S)}$ & $\mathrm {  \log\varepsilon(Cr)}$ & $\mathrm {  \log\varepsilon(Ni)}$ & $\mathrm {  \log\varepsilon(Fe)}$\\
  \noalign{\smallskip}
     &  &  [$\mathrm{dex}$] & [$\mathrm{dex}$] &  [$\mathrm{dex}$] & [$\mathrm{dex}$] & [$\mathrm{dex}$]  & [$\mathrm{dex}$]  & [$\mathrm{dex}$]  & [$\mathrm{dex}$]  & [$\mathrm{dex}$] \\
  \noalign{\smallskip}
  \hline
  \noalign{\smallskip}
 \textrm {Arcturus}&   this work & 8.06 & 7.67 &  8.76 & 4.68 & 7.35 & 6.94 & 5.17 & 5.78 & 7.00  \\
 \noalign{\smallskip}
  & \citeauthor{fulbright:07} &  - & - & 8.67  & 4.68 & 7.39 & - &- & - & 6.95 \\
   \noalign{\smallskip}
    & difference &  -& -& 0.09 & 0.00 & $-$0.04  & - & - & -& 0.05 \\ 
 \noalign{\bigskip}
 \textrm {Arp\,4203} & this work &  6.62 & 7.70 & 7.71  & 3.98 & 6.75 & 6.26 & 4.28 & 5.12 & 6.25 \\
 \noalign{\smallskip}
  & \citeauthor{fulbright:07} &  - & - & 7.55  & 4.03 & 6.82 & - & -& - & 6.20\\  
   \noalign{\smallskip}
    & difference &  -& -&  0.16  & -0.05 & $-$0.07  & - & - & - & 0.05\\
 \noalign{\bigskip}
 \textrm {Arp\,4329} & this work &  7.42 & 7.23 & 8.25 & 4.33 & 7.15 & 6.81 &  4.77 & 5.41 & 6.60\\
 \noalign{\smallskip}
  & \citeauthor{fulbright:07} &  - & - & 8.16  & 4.30 & 7.14 & - & - & - & 6.55\\
   \noalign{\smallskip}
    & difference &  -& -& 0.09  & 0.02 & 0.01  & - & - & -& 0.05\\
\noalign{\bigskip}
   \textrm {Arp\,1322} & this work &  7.93 & 8.10 &8.61  & 4.87 & 7.52 &  7.35 & 5.62 & 6.02 & 7.33  \\
 \noalign{\smallskip}
  & \citeauthor{fulbright:07} &  - & - & 8.80   & 4.84 &7.43 & - & - & - & 7.22\\
   \noalign{\smallskip}
    & difference &  -& -& $-$0.19 &0.03 &0.09 & - & - & - & 0.11 \\
\noalign{\smallskip}
    \hline
  \end{tabular}
\begin{list}{}{}
\item[$^{\mathrm{a}}$]   $\mathrm { \log\varepsilon(X)}=\log n_\mathrm X/n_\mathrm H + 12$, where $\log n_\mathrm X$ is the number density of element X.
\end{list}
\end{table*}

\section{Analysis}

The spectral resolution is high enough for spectral lines to be resolved. Furthermore, in the near-IR wavelength region there are few enough lines that many lines are not blended.  This makes the abundance analysis easier, since individual, unblended lines can be studied.  In the spectra observed we have identified approximately 50 CN, 40 OH, and 40 CO lines of varied quality for an abundance analysis. We determined the carbon abundance from the CO($v=3-0$) band, the nitrogen abundance from a dozen suitable CN lines, and the oxygen abundance from a dozen suitable OH lines. Furthermore, many
metal lines are identified and used in the analysis.

We  analyse our data by modelling the stellar atmospheres of our observed stars and
calculating synthetic spectra, using extensive line lists, for that atmosphere.  The temperature-dependent partition functions and continuous opacities are calculated for every depth in the atmosphere. The spectral region we observed and synthesise ranges from $\lambda_\mathrm{air} = 15315$ to $15710\,\AA$. The synthetic spectrum is thereafter convolved
with a macroturbulence function in order to fit the shapes and
widths of the lines, including the stellar macroturbulence and instrumental broadening (given by the spectral resolution). We then derive elemental abundances by fitting the synthetic to the
observed spectra by changing the atomic abundances, which are used in the calculation of the synthetic spectra and the molecular equilibria.  Finally, we reiterate the method using the derived abundances in the model atmosphere calculation in order to be self-consistent. 

The best fits are found by visual inspection line by line, interpolating in three  synthetic  spectra with incremental differences in C, N, or O abundance of 0.05 dex. For every suitable OH and CN line the best fit is determined. For the CO band a best general fit to the feature is determined. 
In Figure \ref{CNO-uncert} we demonstrate how clearly a difference of $0.1$~dex in the abundances of C, N, or O from the observed spectra can be determined. Note, that for our bulge stars, the CO lines are relatively independent of the O and N abundances and that the OH lines are relatively independent of the C and N abundances. In contrast, the CN lines get stronger with increasing C and get weaker with increasing O. However, with well defined C and O abundances from many lines, the CN lines are mainly dependent on the N abundance.  Using Arp4329 as an example, we find the following abundances and standard deviations: {\it (i)} $\log\epsilon(C)=7.42\pm 0.05$ from the CO band, {\it (ii)}  from the many OH we find $\log\epsilon(O)=8.25\pm 0.03$, and  {\it (iii)}  from the many  CN lines $\log\epsilon(N)=7.23\pm 0.05$, the latter two with a standard deviation of the mean of $0.01$. Hence, we find a typical standard deviation of 0.05 dex for the observational uncertainties.

Next, we will discuss the model
atmospheres, the stellar parameters,  the spectrum synthesis, and the
line data.

\subsection{Model atmospheres}

\begin{table}
  \caption{Stellar parameters from \citet{fulbright:06,fulbright:07}.}
  \label{par}
  \begin{tabular}{r c c c  c c}
  \hline
  \noalign{\smallskip}
    Star & $T_\textrm{eff}$ & $\log g$  &  [Fe/H] & $\xi_\textrm{micro}$ & [$\alpha$/Fe]\\
      & [K] & cgs &  & [km\,s$^{-1}$]  &  \\
  \noalign{\smallskip}
  \hline
  \noalign{\smallskip}
 \textrm {Arp\,4203} & 3902  & 0.51 &  $-$1.25 &  1.9  & 0.4 \\
  \noalign{\smallskip}
 \textrm {Arp\,4329} &  4197  & 1.29 & $-$0.90 & 1.5 & 0.4 \\
  \noalign{\smallskip}
 \textrm {Arp\,1322} & 4106   & 0.89   & $-$0.23 & 1.6 & 0.2\\
 \noalign{\smallskip}
  \hline
  \end{tabular}
\end{table}

We use model atmospheres provided by  the MARCS code \citep{marcs:08}. 
These hydrostatic, spherical model photospheres are
computed on the assumptions of local thermodynamic equilibrium
(LTE), chemical equilibrium, homogeneous spherically-symmetric stratification, and the
conservation of the total flux (radiative plus convective; the
convective flux being computed using the mixing length
recipe). The  radiation field used in the model generation is
calculated with absorption from atoms and molecules by opacity
sampling at approximately 95\,000
wavelength points over the wavelength range $1300\,\mbox{\AA} $--$
20\,\mbox{$\mu$m}$. The models are calculated with 56 depth points from
a Rosseland optical depth of $\log \tau_\mathrm{Ross}=2.0$ out to $\log
\tau_\mathrm{Ross}=-5.0$.
Data on absorption by atomic species are collected from the
VALD database \citep{vald} and Kurucz and other authors (for details, see \citet{marcs:08}). The opacity of CO, CN, CH, OH, NH, TiO, VO, ZrO,
H$_2$O, FeH, CaH, C$_2$, MgH, SiH, and SiO are included and
up-to-date dissociation energies and partition functions are used.

The fundamental parameters of the star are needed
as input for the model photosphere calculation. Based on optical spectra and an extensive discussion,  \citet{fulbright:06,fulbright:07} derived these parameters for, among others, our 3 programme stars.  In our analysis we use these stellar parameters, see Table \ref{par}.  \citet{fulbright:06} find a good agreement for these stars between the Kurucz models they use and MARCS models. Our models are calculated in spherical geometry, assuming a mass of $0.8\,M_\odot$.  Unfortunately, the range in excitation energy for the molecular transitions in our wavelength region is not
great enough to admit a good determination of effective temperature from the spectra alone. In future
studies, the possibilities to extend the wavelength region in order to admit such determinations should be explored.

The iron abundances of  \citet{fulbright:07} are used in the models. In the model atmosphere calculation we have used a general [$\alpha\mathrm{/Fe]} = 0.4$ (including oxygen) for the two most metal-poor stars, whereas for Arp\,1322 we use [$\alpha\mathrm{/Fe]} = 0.2$ (based on the discussion by Melendez et al. 2009, in prep.). We also use specific abundances for several elements. {\it (i)} The  C, N, O, Si, S, Ti, Cr, and Ni abundances are derived from our spectra. {\it (ii)} The logarithmic Mg, Al, and Na abundances are taken from the analysis by \citet{fulbright:07} for these stars, which are \{7.44, 6.34, 5.89\}, respectively for Arcturus, \{6.74, 5.63, 5.24\} for Arp\,4203, \{7.06, 5.85, 5.56\} for Arp\,4329, and \{7.52, 6.65, 6.36\} for Arp\,1322, on the usual scale where the hydrogen abundance is set to 12.  Among these abundances Mg is important as an electron donor and may therefore affect the equivalent widths of spectral lines through the continuous opacity, which is mainly due to H$^-$, while Al and Na are less significant. In a model of, for instance Arp\,1322, the most important electron donors at $\tau_\mathrm{Ross}\sim 1$ are, in order of importance, Mg, Si, Fe, Al, H, Na, and  Ca. The further out in the atmosphere, the relatively more important Al
and Na become. Mg and Fe are, however, still the most important contributors in shallower layers. 

The atomic opacity files used in a MARCS calculation are pre-calculated in a grid. The ones used for our stars are thus files with a general metallicity of [Fe/H] =  $-1.25$, $-1.00$, and $-0.25$, for Arp\,4203, Arp\,4329, and Arp\,1322, respectively. The [$\alpha$/Fe] enhancement in the pre-tabulated opacity files are 0.4, 0.4, and 0.2, respectively. Furthermore, the opacity files are calculated with  a microturbulence parameter of $2\,\mathrm{km\,s^{-1}}$. 

\subsection{Synthetic spectra}

For the purpose of analyzing our observations, we have generated
synthetic spectra, also calculated in spherical geometry, based on these model photospheres.   Calculating a synthetic spectrum from a given line list and given abundances, is done with the program BSYN v. 7.06, and  calculating an abundance for an element from measured equivalent widths, is done with the program  EQWI  v. 7.05. These programs are based on procedures from the MARCS code. 

The same general metallicity ([Fe/H]), [$\alpha$/Fe], and individual abundances of the specified elements, are used as in the calculation of the model atmosphere. In addition a $^{12}\mathrm C/^{13}\mathrm C = 24$ (96\% $^{12}\mathrm C$) is used for the bulge stars, and $^{12}\mathrm C/^{13}\mathrm C = 9$ for $\alpha$ Boo. For elements which are not  trace elements, but whose abundances affect the model structure, we iteratively change the abundance  in the calculation of the model atmospheres, in order to be self-consistent. 

We have synthesised the entire observed wavelength range. 
We calculate the radiative transfer for points in the spectrum 
separated by $\sim 0.5\,\mathrm {km\,s}^{-1}$ (corresponding to a resolution of 
$R=600,000$), although the final resolution is lower. With microturbulence velocities of $1.5\,\mathrm{km\,s}^{-1}$ or more, this will ensure an adequate sampling in the generation of the synthetic 
spectra.

Finally, to match the observed line profiles, we introduce the customary artifice of a macroturbulent broadening, with which we 
convolve our synthetic spectra with a radial-tangential 
function \citep{gray:1992}.  This extra broadening also includes the instrumental profile and does not 
change the equivalent widths of the lines. We find that we need a macroturbulence of $\xi_{\mathrm{macro}}= 6.2$, $5.8$, and $6.2\,\mathrm{km}\,\mathrm s^{-1}$ (FWHM) for Arp\,4203, Arp4329, and Arp1322, respectively.

\subsubsection{Atomic and molecular line lists}

The line lists needed for the generation of the synthetic spectra consist of tables of wavelengths, excitation energies of the lower 
state of the transition, line strengths in the form of oscillator strengths, and  the statistical weight of the upper level of the transition for molecules and atoms. Depending  on the transition, the atomic line lists also provide the collisional line-broadening  computed according to the collisional broadening theory by Anstee \& O'Mara, \citep[see for instance][]{anstee,barklem:00,aspelund} or  the damping enhancement factor for van der Waals broadening. Furthermore, the lists provide  the radiative damping ($\Gamma_\mathrm{rad}$), and the electronic orbitals and designation of the levels involved in the transition.

The atomic line list is compiled from the VALD database \citep{vald}. In addition, we have determined `astrophysical gf-values' by fitting atomic lines in synthetic spectra to the solar spectrum \citep{solar_IR_atlas}, see Figure \ref{sun} and Table \ref{linelist}.  The lines fitted were, among others, some Fe, Ni, S and Ti lines.  

The lines which are too weak in the solar spectrum but visible in the $\alpha$ Boo spectrum are fitted to the `summer' version
of the $\alpha$ Boo\footnote{The fundamental parameters of Arcturus are closer to those of our programme stars} IR atlas of \citet{arcturus}, see Figure \ref{aboo} and Table \ref{linelist}.  In principle, there are, in the wavelength range observed, 11 Si lines (1 saturated), 7 Ti (2 saturated), 4 S, 5 Ni, 1 Cr, and many Fe lines to be used for an abundance analysis in bulge giants similar to ours.  A few lines are avoided, such as the Ti lines at $\lambda15334.8$ and $15543.8$~\AA\  ($W/\lambda\sim-4.9$) which  are saturated and therefore less sensitive to the abundance and quite sensitive to the microturbulence. The parameters of our 
$\alpha$ Boo model are adopted as in \citet{fulbright:07}, namely, 
$T_\mathrm{eff}=4290$~K, $\log g = 1.55$ (cgs), [Fe/H]$=-0.50$, [$\alpha$/Fe]$=0.4$, and $\xi_\mathrm{micro}=1.7$. The abundances of $\alpha$ Boo of Mg, Al, Na, Si, Ti, and Ca are taken from \citet{fulbright:07}. The synthetic spectra are convolved with a macroturbulence parameter  of $4$~km\,s$^{-1}$  for
the comparison with the atlas.
It would appear that the atlas was normalised such that the very wide
Bracket hydrogen lines were removed. Therefore these were  not included
in the synthetic spectrum
calculations, which will have some effects on the line widths.

%

The molecular line lists are given for all isotopic combinations that are relevant. 
The molecular lists included are, for CO  \citep{goor},  SiO (Langhoff \& Bauschlicher)\nocite{lang}, CH \citep{jorg}, CN (J\o rgensen \& Larsson, 1990; Plez, 1998, private communications)\nocite{jorg_CN}, OH \citep{gold}, and C$_2$ (Querci et al. 1971; J\o rgensen, 2001, private communications)\nocite{querci}. The accuracy and the completeness of these line  lists are discussed in \citet{leen_phd}.
For the molecules, the line lists were adopted as they are and instead
of modifying the $gf$ values, 
the abundances of $\log\epsilon_\mathrm O =8.76$ (from OH lines), then $\log\epsilon_\mathrm C =8.06$ (from CO lines) and
last $\log\epsilon_\mathrm N=7.67$ (from CN lines) were obtained from the Arcturus atlas. These values are in fair agreement with the values found by \citet{leen_aBoo} and used in \citet{ryde:02}, namely  $\log\epsilon_\mathrm O =8.67$,  $\log\epsilon_\mathrm C =7.90$ $\log\epsilon_\mathrm N=7.55$.  The OH and CO lines are not
seen in the Solar atlas and can therefore not be checked. In contrast to the case for the OH and CO lines, the wavelengths of the CN lines very often seem to be wrong,  giving rise either to too strong or too weak features, and
sometimes features where there are no observed lines at all.
Therefore the judgement of the general fit for the CN lines will be uncertain.
A few obvious shifts were made. The CN lines in the
Sun seem to become too weak when our C, N, and O abundances of 8.41, 7.80,
and 8.66, respectively  \citep{solar_asplund}, are assumed. This improves if instead the
\citet{solar2} abundances are used. This might perhaps be expected considering
the 3D modelling origin of former abundances and the 1D plane-parallel models
used here.  






\begin{table*}
  \caption{Abundances  for Arcturus and the three bulge giants relative to the Sun, where $\mathrm{[X/Fe]} = \{\mathrm { \log\varepsilon(X)} - \mathrm { \log\varepsilon(Fe)\}_{star}} - \{\mathrm { \log\varepsilon(X) } - \mathrm { \log\varepsilon(Fe)\}_\odot }$. }
  \label{abundances2}
  \begin{tabular}{l l c c c c c c c c c c c}
  \hline\hline 
  \noalign{\smallskip}
    Star &  Ref.   Star &  [C/Fe]   & [N/Fe] & [C+N/Fe] &  [O/Fe]  & [Ti/Fe]& [Si/Fe] & [S/Fe] &[Cr/Fe] & [Ni/Fe] & [Fe/H]\\
  \noalign{\smallskip}
  \hline
  \noalign{\smallskip}
 \textrm {Arcturus}&   this work$^{\mathrm{a}}$ & 0.15 & 0.37 & 0.20  & 0.60  & 0.16 & 0.30 & 0.24 & 0.00 &0.03  & -0.50\   \\
 \noalign{\smallskip}
  & \citeauthor{fulbright:07}$^{\mathrm{b}}$ &  - & - & -  & 0.51  & 0.16 & 0.34 & - &- & - & -0.50\\
 \noalign{\bigskip}
 \textrm {Arp\,4203} & this work$^{\mathrm{a}}$ &  $-$0.54 & 1.15 & 0.48 & 0.30  & 0.21 & 0.45 & 0.31 & $-$0.14 & 0.12 & $-$1.25\\
 \noalign{\smallskip}
  & \citeauthor{fulbright:07}$^{\mathrm{b}}$ &  - & - &-  &  0.14  & 0.26 & 0.52 & - & -& - & $-$1.25 \\  
   \noalign{\bigskip}
 \textrm {Arp\,4329} & this work$^{\mathrm{a}}$ & $-$0.09  & 0.33& 0.03 & 0.49  & 0.21 & 0.50 & 0.51 & 0.00  & 0.06 & $-$0.90\\
 \noalign{\smallskip}
  & \citeauthor{fulbright:07}$^{\mathrm{b}}$ &  - & - & -  & 0.40  & 0.18 & 0.49 & - & - & - & $-$0.90 \\
  \noalign{\bigskip}
   \textrm {Arp\,1322} & this work$^{\mathrm{a}}$ &  $-$0.25 & 0.53& 0.05 & 0.18 & 0.08 & 0.20 & 0.38 & 0.18 & 0.00& $-$0.17$^{\mathrm{c}}$\\
 \noalign{\smallskip}
  & \citeauthor{fulbright:07}$^{\mathrm{b}}$ &  - & - & -  &  0.37   & 0.05 & 0.11 & - & - & - & $-$0.23\\
\noalign{\smallskip}
    \hline
  \end{tabular}
\begin{list}{}{}
\item[$^{\mathrm{a}}$] We have used the following solar abundances:  $\mathrm { \log\varepsilon(C)=8.41}$, $\mathrm { \log\varepsilon(N)=7.80}$, $\mathrm { \log\varepsilon(O)=8.66}$, $\mathrm { \log\varepsilon(Fe)=7.50}$, $\mathrm { \log\varepsilon(Ti)=5.02}$, $\mathrm { \log\varepsilon(Si)=7.55 }$, $\mathrm { \log\varepsilon(S)=7.20}$, $\mathrm { \log\varepsilon(Cr)= 5.67}$, and $\mathrm { \log\varepsilon(Ni)=6.25}$.
\item[$^{\mathrm{b}}$]   \citeauthor{fulbright:07} used the following solar abundances:  $\mathrm { \log\varepsilon(O)=8.69}$, $\mathrm { \log\varepsilon(Fe)=7.45}$, $\mathrm { \log\varepsilon(Ti)=4.92}$, and $\mathrm { \log\varepsilon(Si)=7.54}$. For the sake of comparison we have scaled their values to the solar values we have used.
\item[$^{\mathrm{c}}$]   The [Fe/H] from this work  is determined from Fe{\sc i} lines in the CRIRES wavelength region, but not used as the metallicity in the models.
\end{list}
\end{table*}

  \begin{figure*}
   \centering
 \includegraphics[ width=\textwidth]{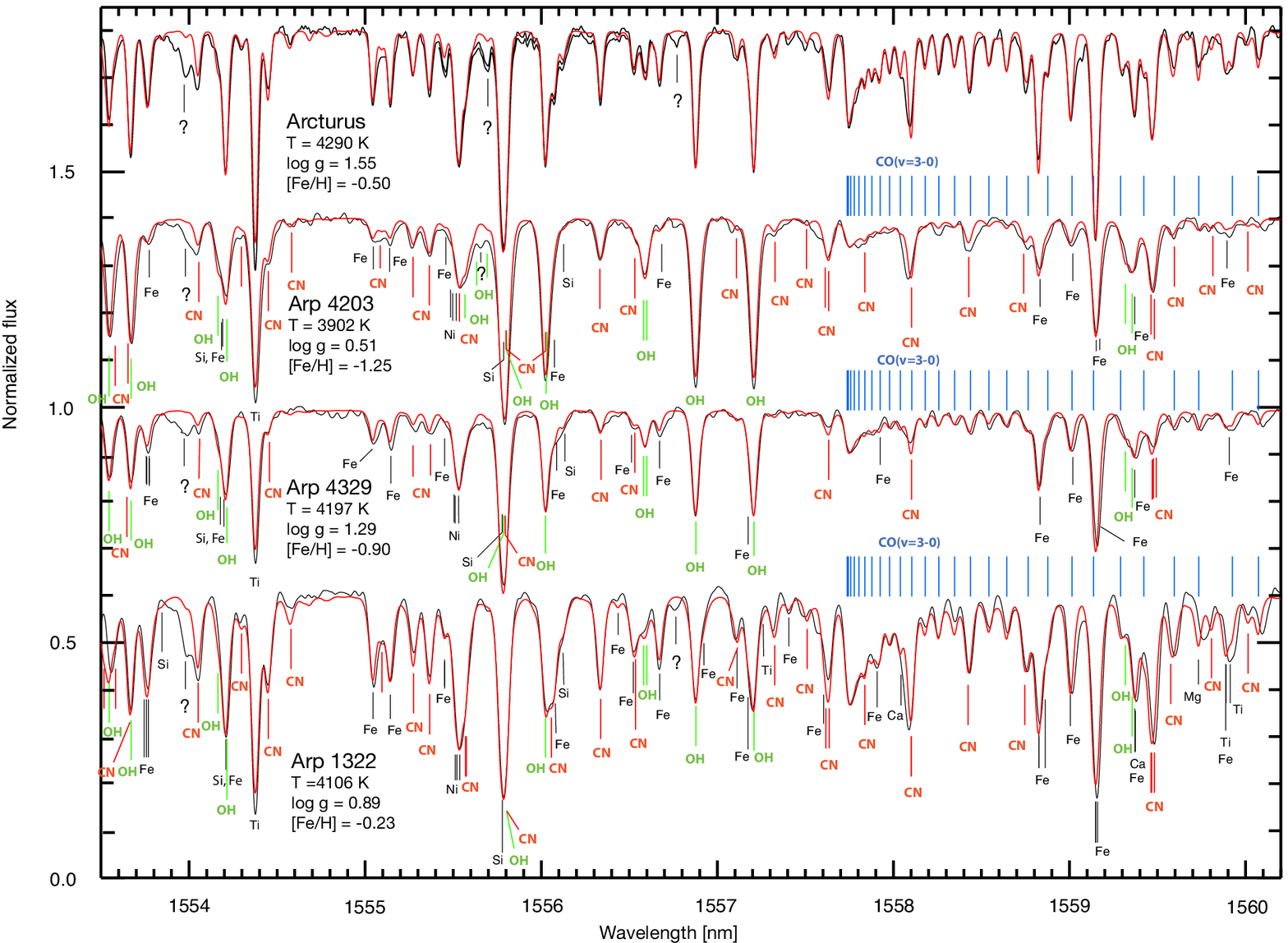}
      \caption{Sections of the observed CRIRES spectra of the three bulge giants Arp\,4329, Arp\,4203, and Arp\,1322 in Baade's Window are shown with full, black lines. For comparison, the Arcturus atlas spectrum \citep{arcturusatlas_II} is also shown. The parts of the spectra which have the highest signal-to-noise ratios are shown. In these parts  the C, N, and O elements can be determined. Our best synthetic spectra is shown in red. All synthetic lines which are deeper than 0.97 of the continuum are marked. A few features are not identified in the Arcturus spectrum and are labelled with question marks. These features also show up in the bulge-star spectra. }
         \label{obs_model}
   \end{figure*}
\section{Discussion}

In Figure \ref{obs_model} we show our observations of the three observed bulge giants and the Arcturus spectrum, together with our final synthetic spectra. Only data from the detector array with the highest SNR are shown here (the third one of four). The bulge-star spectra are
shown in the order of increasing metallicity and CO-band strength. The fits are, in general, very good. It is obvious that there are numerous, clean molecular-lines of CO, CN, and OH available in this region from which  the  C, N, and O abundances, as well as the important C-N-O molecular equilibria, can be determined. This should be compared with the scarcity of relevant lines in the optical region.  

 Since the uncertainties in the abundances derived from molecular lines due to uncertainties in the fundamental parameters are 
of importance, we append a discussion of the sensitivities of the derived abundances to the fundamental parameters in Appendix A.

\subsection{Uncertainties in the derived C, N, and O abundances}

 For a typical uncertainty of $+100$\,K, for a model at 4000 K, uncertainties  of $\Delta A_\mathrm C = +0.05$\,dex,  $\Delta A_\mathrm N = +0.06$\,dex, and  $\Delta A_\mathrm O = +0.14$\,dex are found.
A typical uncertainty in $\log g$ of $+0.2$~dex, gives correspondingly, also at $T_\mathrm{eff}=4000$\,K,  $\Delta A_\mathrm C = +0.08$\,dex,  $\Delta A_\mathrm N = -0.02$\,dex, and  $\Delta A_\mathrm O = +0.03$\,dex. Similarly, for an uncertainty in the microturbulence of $+0.5\,\mathrm{km\,s}^{-1}$, one gets  $\Delta A_\mathrm C = -0.02$\,dex,  $\Delta A_\mathrm N =-0.07$\,dex, and  $\Delta A_\mathrm O = -0.03$\,dex. Low values are, indeed, expected for weak lines, such as these typical lines used in this test.  Finally, uncertainties in the metallicity and [$\alpha$/Fe] of $+0.1$~dex yield uncertainties in the C, N, and O abundances of $+0.08$~dex and $+0.02$~dex, respectively. Thus, we estimate the total internal uncertainties in the derived C, N, and O abundances, assuming uncertainties in the stellar parameters to be uncorrelated, 
to be approximately $\Delta A_\mathrm C = \sqrt{0.05^2 + 0.08^2 + 0.02^2+0.08^2+0.02^2} = 0.13$,  $\Delta A_\mathrm N = \sqrt{0.06^2 + 0.02^2 + 0.07^2+0.08^2+0.02^2} = 0.13$, and  $\Delta A_\mathrm O = \sqrt{0.14^2 + 0.03^2 + 0.03^2+0.08^2+0.02^2} = 0.17$. As a comparison, the standard deviations in the determinations of the C, N, and O abundances from the many observed CO, CN, and OH lines for a given model are small, less than 0.05 dex.

A source of uncertainty for N abundances might also be the dissociation of energy of CN, which
is here, following \citet{costes} set to $\mathrm{D}_0= 7.77$~eV. This quantity used to
be highly uncertain. We note, however, that \citet{pradhan}  found the value
$\mathrm{D}_0=7.72 \pm 0.04$ from multireference configuration-interaction calculations, which also agrees with  experimental results of \citet{huang}  of $7.74 \pm 0.02$. Thus, at least differentially
to Arcturus, the errors in N abundances due to this should be negligible.

 Whereas both weak and strong OH lines are fitted nicely in Arp\,4329 and Arp\,1322, the strong ones are too weak in the model of Arp\,4203 and the weak ones too strong. The same is found for the Arcturus spectrum. This problem can not be solved by the modification of effective temperatures, and is thus not understood but will be investigated later.

\subsection{Abundances}

A part of the analysis is presented in \citet{ryde:07}, where we provide our preliminarily derived elemental abundances. Our new resulting abundances of C, N, O, Ti, Si, S, Cr, Ni, and Fe 
are presented in Tables \ref{abundances1} and  \ref{abundances2}.  As a comparison the  abundances for the same stars determined by  \citet{fulbright:07,fulbright:06} are also provided in the Tables.
Typical uncertainties for the abundances derived from atomic lines 
are of the order of $0.1$~dex and of the order of $0.15$~dex from molecular lines.  The  uncertainties in the derived oxygen abundances by Fulbright et al. (2007) are approximately $0.15$~dex (from the [OI] $\lambda 6300$ line) and $0.1$~dex for Si and Ti. A good agreement is found
between our near-IR abundances of O, Ti, Si, and Fe and the abundances from Fulbright et al., 
an analysis based on optical spectra. The largest discrepancy, albeit within the uncertainties,  is in the oxygen abundance, which is not surprising. The good agreement is reassuring for future analysis of stars for which only near-IR spectra will exist, for instance, analyses of bulge stars closer to the Galactic plane. It should be noted, however, that a general problem in determining elemental abundances for stars with only near-IR spectra is the temperature sensitivity of the molecular lines and the difficulties in the determination of the stellar parameters. 

In Table \ref{abundances2} the relative abundances of the elements for Arcturus and the three bulge giants relative to the solar values are given. We see that Arcturus has quite a large over-abundance of [O/Fe] which
is approximately three times that expected for disk stars at [Fe/H]$=-0.5$. The [O/Fe] for the three bulge stars fits very well with the trend discussed in \citet{melendez:2008}. Our two `metal-poor' stars lie on the plateau  defined by both disk and halo stars.  Our determination of the [O/Fe] in Arcturus lies at the high end, mostly populated by thick disk stars \citep[see Figure 2 in][]{melendez:2008}. Our low [O/Fe] value for Arp\,1322 with [Fe/H]$=-0.25$ corroborates the finding of   \citet{melendez:2008} that there is no obvious difference between the oxygen abundance in bulge stars and those in thick disk stars. More stars are, however, needed in order to confirm this trend.  Note that in a recent paper \citet{chiappini:08} compare, among others, the oxygen abundances derived from planetary nebulae (PNe) and giants in the bulge and find an interesting discrepancy that the abundances determined from  giant star spectra are systematically higher by 0.3 dex. They conclude that this discrepancy may be caused by systematic uncertainties in either the PNe or giant star abundance determinations, or both.

The carbon abundances are all depleted in our bulge giants, and  nitrogen is enhanced. In Arp\,4329 and Arp\,1322
the C+N seems, however, conserved. This is, indeed, expected for giants that have ascended the giant branch for the first time. They have experienced CN-cycling in their interiors (with $^{14}\mathrm{N}$ and $^{13}\mathrm{C}$ as products) 
and experienced the first dredge-up. The giant Arp\,4203 shows a large depletion of carbon and a large enhancement of nitrogen. Further, the [C+N/Fe] is non-solar, making this giant special. \citet{fulbright:07} also note the peculiarity of this giant. They argue that it is a cluster member of NGC 6522 based on their oxygen, high sodium, and aluminum abundance pattern. A possibility is that this giant has experienced nuclear processing  in 
H-burning shells, with subsequent dredge-up of  material to the surface. This burning would also be 
responsible for the O-Na correlations in globular clusters. The high [C+N/Fe] we find may fit with this picture, since
one expects an increased [N/Fe] ratio due to CNO-cycling \citep{gratton:04,melendez:2008}. 
We also find that Arp\,4203 has an excess of O relative to Fe ($\mathrm{[O/Fe]}=0.3$), which may, however, be slightly lower than the assumed general $\alpha$-element enhancement of $+0.4$. The sum of the abundances of C, N, and O ([C+N+O/Fe]$\sim0.4$) is close to that expected from a non-processed old star, formed with an excess of oxygen and $\alpha$ elements relative to iron. Thus, if the carbon and nitrogen abundances are correct, this implies that the oxygen abundance in this star should actually not be used to represent the unprocessed value for this bulge giant.

As regards the $\alpha$ elements Si and Ti we find abundance values which agree with those of \citet{fulbright:07}. Our sulphur abundances 
are the first determinations for bulge stars as
far as we know. All [S/Fe] ratios are relatively high, also for our most metal-rich star ([S/Fe]$=+0.4$ at [Fe/H]$=-0.2$). This implies a high star-formation rate in an early phase of the bulge evolution as discussed in the Introduction.

The Cr abundance is based on one line which introduces uncertainties due to the continuum fitting and a CN line blending into the line. This is true in particular for the most metal-rich star of our sample. Thus, given the uncertainties, 
we cannot say whether the Cr abundance follows the general metallicity or not. This is also true for our abundance result for Ni.


It is interesting to note that lines from Si, S, Ni, and Cr in $\alpha$ Boo are nearly of the same strength as in the Sun. The Sun is more metal-rich and hotter which would increase the strength of these highly excited lines ($\chi_\mathrm{exc}=5-8$~eV). However, the continuous opacity due to H$^{-}$ is also larger, which thus compensates for this effect.
Some Fe lines are stronger, others slightly weaker, and the Ti lines are much stronger in Arcturus.

\section{{\bf Conclusions and outlook}}

Stellar surface abundances in bulge stars can be extensively studied in the near-IR, due to lower extinction. 
In particular, the abundances of the C, N, and O elements are advantageously studied in the near-IR compared to the optical wavelength region, due to numerous CO, CN, and OH lines.  All these molecules are observable in the same part of the spectrum and can therefore be studied simultaneously in order to determine the molecular equilibria properly. Here,  we present the first three CRIRES spectra of bulge stars, observed during the {\it science verification observations} of the CRIRES spectrometer at the {\it VLT}. We show that we can determine the C, N, and O abundances from molecular lines, with uncertainties of $\sim 0.15$~dex, which are mainly due to uncertainties in the stellar parameters.  
Unfortunately, the range in excitation energy for the molecular lines is not great enough for determinations of effective temperatures from our spectra. The possibilities to extend the wavelength region in order to admit such determinations should be explored.

 For oxygen, titanium, iron, and silicon, we show a good agreement between near-IR and optically determined abundances in stars in Baade's window, stars which can be observed in both wavelength ranges. For the two of our stars that show an unprocessed oxygen abundance, the [O/Fe] trend corroborates the idea that there is no
significant difference between the oxygen abundances in bulge and thick disk stars. The abundance of the $\alpha$ element sulphur indicates a high star-formation rate in an early phase of the bulge evolution.

It will be very important to extend the analysis to more stars and especially to stars in 
other regions of the Galactic bulge, such as in the Galactic plane, in order to get a proper handle on the formation and evolution of the bulge. In these regions the optical extinction is high which permits observations only in the near-IR. Near-IR, high-spectral-resolution spectroscopy offers a promising methodology to study the whole bulge to give clues to its formation and evolution. 

The observations presented here have demonstrated the feasibility and
power of precision abundance measurements in the H-band. Some technical
shortcomings of these early science verification observations have
triggered the investigation into better non-linear flat fielding.
Moreover this work has triggered the search for ways to improve
throughput and efficiency.  It has therefore been important for defining the
scientific requirements and the scope of the CRIRES upgrade plan, and in
this context changing the Echelle format by procuring a custom-ruled
grating. Then it will be possible to use all pixels in the spectrograph
focal plane without compromises.

\begin{acknowledgements}
The anonymous referee is thanked for valuable comments. We would also like to thank the CRIRES science verification team for their work on CRIRES and for the execution of the observations. NR is a Royal Swedish Academy of Sciences research fellow supported by a grant from the Knut and 
Alice Wallenberg Foundation. Funds from Kungl. Fysiografiska S\"allskapet i Lund are acknowledged.
BE, BG, and KE acknowledge support from  the Swedish research council, VR. Karin Ryde is thanked for 
reading through the manuscript and correcting the English.

\end{acknowledgements}

\section{Appendix A. Sensitivities in the derived abundances to the stellar parameters}

\begin{table}
\caption{Test lines}             
\label{table:lines}      
\centering                          
\begin{tabular}{l c c c c c}        
\hline\hline                 
Molecule & $Wavelength$ & $\chi_\mathrm{exc}$ & $\log gf$ & $W$ & $\log W/\lambda$ \\    
 & [\AA] & [eV] & (cgs) & m\AA\ \\
\hline                        
   CO &  15580.4 &  0.37 & $-7.44$ &  60 &  $-5.4$\\      
   CN &  15563.4 &  1.15 & $-1.14$ & 120 & $-5.2$\\
   OH  & 15568.8  & 0.30 & $-5.29$ &  152 & $-5.0$  \\
   \hline                                   
\end{tabular}
\end{table}
 
\label{eqwimo}

In order to test how the derived abundances from a molecular line with a given equivalent width changes
for uncertainties in the fundamental parameters of  a star, a program EQWIMO v.1.0 was written, based on
the MARCS code. The code needs molecular line lists and information on which atom in a molecule the abundance is to be solved for.
Thus, an estimate of the sensitivities of the derived carbon, nitrogen, and oxygen abundances from the CO, CN, and OH lines to the fundamental stellar parameters can be obtained by analyzing  typical molecular lines of specific strengths and deriving the abundances they would yield for different stellar parameters. For this exercise we have assumed that the  stellar parameters are uncorrelated.  The C, N, and O abundances are determined simultaneously for a change in the parameters, by iteratively determining one at a time. The program  recalculates the molecular equilibria for every iteration. Only a few iterations are needed to reach convergence. Apart from being affected by the temperature change, the abundances derived from the molecular lines are also affected by the changes in the C, N, and O abundances through the molecular equilibria.  This is especially the case for the derived nitrogen abundance. In the models that we have generated for this exercise, we have used  $\log \epsilon_\mathrm C  = 8.32$,   $\log \epsilon_\mathrm N  = 8.00$, and $\log \epsilon_\mathrm O  = 8.81$. For many of the models, the derived C, N, and O abundances will be different, but this difference should not be significant for our abundance results. We have analysed  typical CO, CN, and CO lines of the following strengths (equivalent widths): $W_\mathrm{CO}=60\,\mathrm{m\AA}$,$W_\mathrm{CN}=120\,\mathrm{m\AA}$, and $W_\mathrm{OH}=152\,\mathrm{m\AA}$, see Table \ref{table:lines}.  

\begin{figure*}
   \centering
 \includegraphics[angle=90,bb=-10 20 500 750,width=\textwidth,clip]{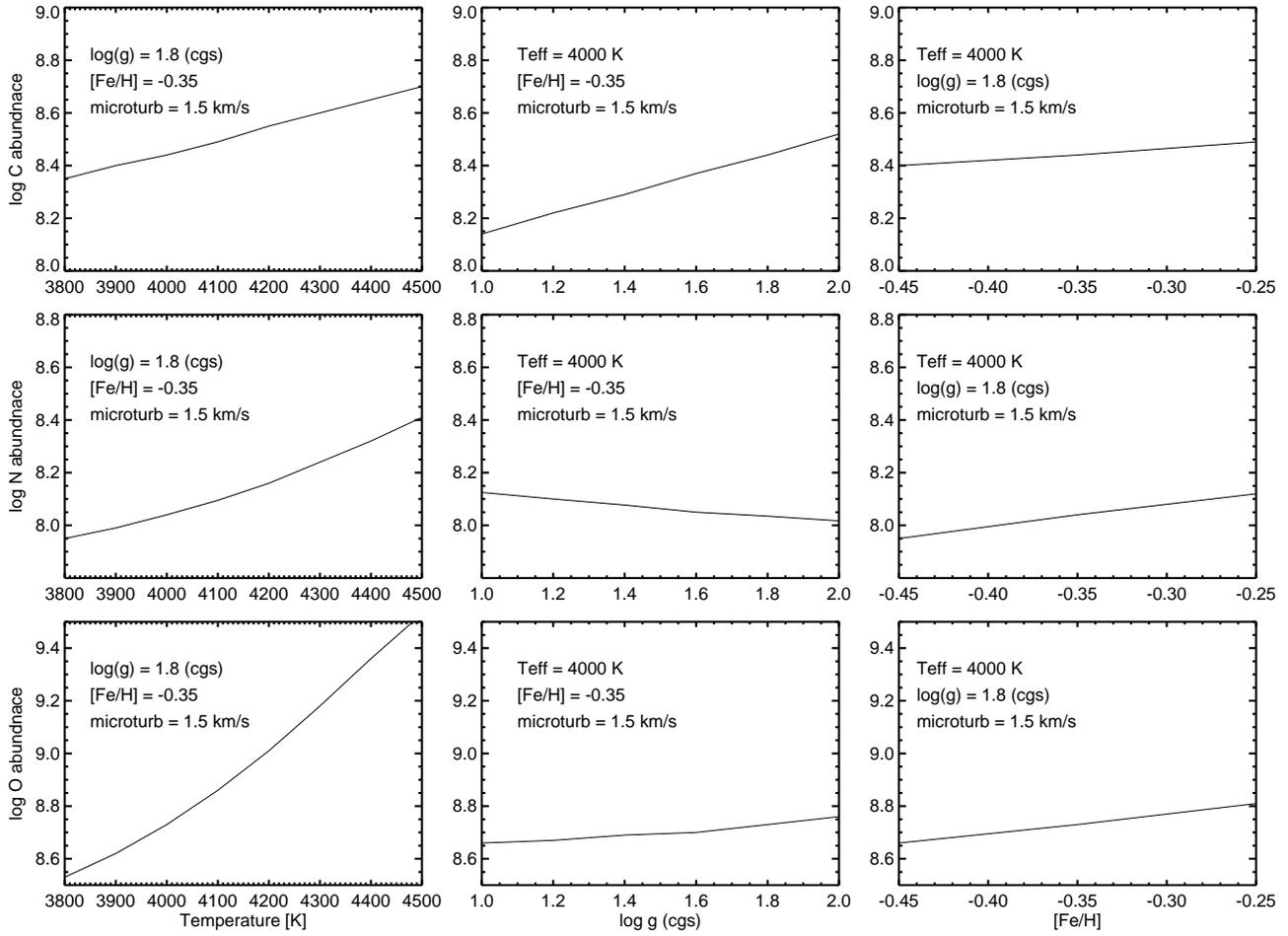}
      \caption{The derived abundances of oxygen, carbon, and nitrogen for different temperatures, surface gravities, and metallicities for  given typical CO, CN, and OH lines at $1.5\,\mu$m of the following strengths (equivalent widths): $W_\mathrm{CO}=60\,\mathrm{m\AA}$,$W_\mathrm{CN}=120\,\mathrm{m\AA}$, and $W_\mathrm{OH}=152\,\mathrm{m\AA}$. To facilitate comparisons, the different panels are in relative ordinate scale,  spanning an order of magnitude in abundance. 
}
         \label{tgxi-uncert}
   \end{figure*}

 In Figure \ref{tgxi-uncert}, the abundances of the C, N, and O elements derived from these typical lines  at $1.5\,\mu$m are shown for different 
effective temperatures, surface gravities, and metallicities, but for given line strengths. We can see that  the oxygen abundance is most temperature sensitive. Both the CO and OH lines are of low excitation and therefore not much affected by the temperature for their excitation. 
The temperature behaviour of the abundances derived from these lines is mainly caused by the increasing dissociation of the molecules (more for OH, which has the lowest dissociation energy of the three molecules) as the temperature rises. Fewer molecules are available at higher temperatures, which means that the abundances of the constituting  atoms have to be higher for a given equivalent width of the molecular line.  The CN line, which has an excitation energy of  its lower level of $1.15$~eV, is more sensitive to the temperature for its excitation. As the temperature increases more CN is excited to the lower level of the CN line, which means that a lower abundance is needed for a given equivalent width. The behaviour of the derived nitrogen abundance from CN is governed by this effect, and by the increased dissociation of the molecule as the temperature increases, but also by the increase of the carbon abundance (which increases the CN abundance) and the large increase of the oxygen abundance (which increases the NO abundance and decreases the CN abundance). In sum, more nitrogen is needed for a given CN line strength as the temperature rises.
 In all these cases the continuous opacity does not change very much.  Since the line strengths are proportional to the ratio of the line to continuous opacities, 
also the continuous opacity has to be discussed when analysing the changes in line strength  due to abundance changes. At $1.5\,\mu$m it is almost solely due to H$^-$ bound-free and free-free processes (more than $98\%$).

From Figure  \ref{tgxi-uncert}, we can also note that the carbon abundance derived from the CO  line  increases with increasing surface gravity, whereas the oxygen abundance derived from the OH line is nearly independent of a change in $\log g$. The reason is that for an increase of the surface gravity from say  $\log g =1.0$ to $\log g =2.0$, the continuous opacity  increases by a factor of 3 (or 0.5 dex). However, also the partial pressure
of OH ($\log P_\mathrm{OH}/P_\mathrm g$) at the formation depth increases by approximately the same amount, which leaves the line strengths nearly unaffected, and thus the oxygen abundances derived from a given equivalent width of the OH line will not vary greatly.  However, the partial pressure of CO  increases by only approximately $60\%$ (0.2 dex). This means that the increase of the continuous opacity 
affects the lines more in the sense that they get weaker as the surface gravity of the stellar model increases. Hence, 
a larger abundance of the constituent atoms is needed for a given equivalent width. The increased carbon abundance increases the CN abundance as the gravity increases, leading to slightly lower nitrogen abundances. Finally, the C, N, and O abundances increase as the metallicity increases, since the continuous opacity increases which weakens the lines. 

\begin{figure*}
   \centering
 \includegraphics[ width=\textwidth]{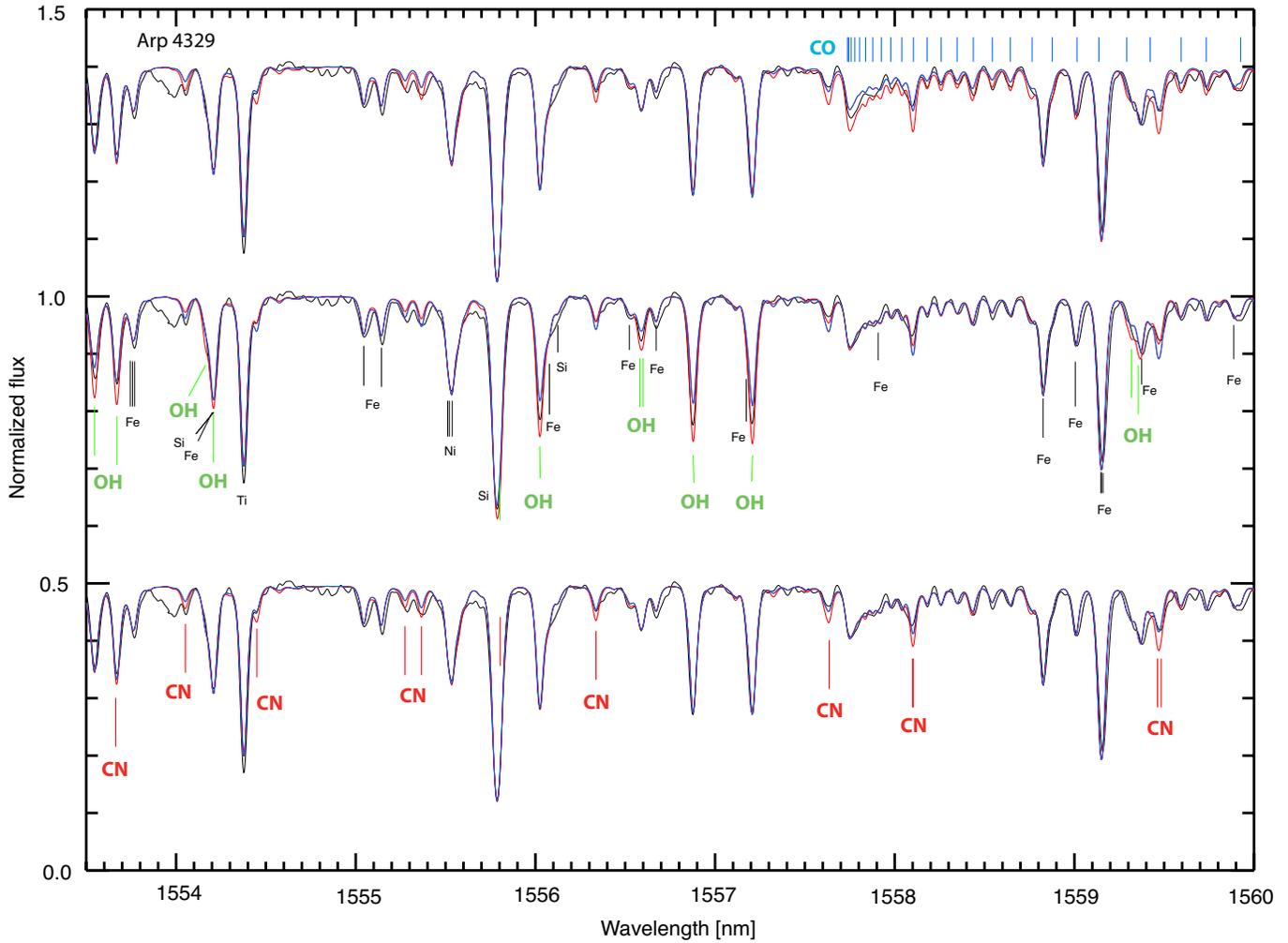}
      \caption{Part of the observed CRIRES spectrum of the bulge giant Arp\,4329 is shown with a full, black line. Synthetic spectra are shown in red and blue, with the carbon, oxygen, and nitrogen abundances, respectively,  changed by $+0.1$ dex (red line) and $-0.1$ dex (blue line) relative to the best-fit abundances.  }
         \label{CNO-uncert}
   \end{figure*}

In Figure \ref{CNO-uncert} we show the spectral changes when changing the C, N, and O abundances by $\pm 0.1$~dex. We have chosen to take our model of Arp\,4329 as a test model. The stellar parameters are given in Table \ref{par}, and, more importantly, the original C, N, and O abundances are those given in Table \ref{abundances1}.  Difference in abundances of this magnitude can clearly be detected in the observed spectra.

Of the three molecules  for which we detect lines, CO is the most abundant, CO being the most stable molecule with the highest dissociation energy  (see Table \ref{table:dis}) and most carbon being locked up in CO. Next in abundance 
comes OH, which is easily dissociated due to a low dissociation energy. Of the nitrogen-bearing molecules, CN is the third most abundant, which makes it a trace molecule. Thus, its abundance-change is to a high degree determined by how the other three molecules, namely N$_2$, NH, and NO, are affected. 


\begin{table}
\caption{Relative logarithmic partial pressures at $\log \tau_\mathrm{Ross}=0$, and dissociation energies of some relevant molecules for a model of 
Arp 4329.}             
\label{table:dis}      
\centering                          
\begin{tabular}{c c c c}        
\hline\hline                 
Molecule & $\log \{P_\mathrm{CO}/P_\mathrm H\}$ & $D^0_0$\\    
 &  at $\log\tau_\mathrm{Ross}=0$ & \\
\hline   
CO &  $-5.8$ &  11.09\\      
   OH & $-7.0$ & 4.39\\
  N$_2$ &  $-8.1$& 9.76\\
  NH & $-8.7$& 3.42\\
 CN & $-9.7$ & 7.76\\
 NO  & $-9.9$ & 6.50 \\
\hline                                   
\end{tabular}
\end{table}

The continuous opacity does not change when the C, N, and O abundances are changed by 0.1 dex. Thus, with the partial pressures of the molecules in mind and noting that oxygen is more abundant than carbon and nitrogen, we can understand qualitatively the changes in Figure \ref{CNO-uncert};
\begin{itemize}
\item Changing the oxygen abundance affects both the OH and CN lines. Increasing the oxygen abundance increases the OH lines and decreases the CN lines by a relatively large amount. The CO lines are actually also increased, but  by a very small amount. The reason for the changes 
is that, in the line-forming regions, the partial pressures increase by a large amount for OH and a small amount for CO, which is to be expected. Also the partial pressure of NO increases, which leads to a smaller amount of available nitrogen. Since also some extra carbon is used in increasing the number of CO molecules, there are smaller amounts of C and N left for the formation of CN, whose line strengths therefore decrease.

\item When changing the nitrogen abundance, only the CN lines are affected discernibly. The partial pressures of N$_2$, NH, and NO are also increased, but the OH and CO lines are not affected noticeably. 

\item As expected, when increasing the carbon abundance both the CO and CN lines get stronger. However,  the OH pressure is not affected very much. In general, determining the oxygen abundance  from OH lines can only be done if the C abundance is also know, since much oxygen is locked up in CO, and a change in the carbon abundance
will affect the amount of available oxygen for OH, see point 5 in Sect. \ref{sectabun}. Indeed, at solar-type C/O ratios, there is only 80\%  more oxygen than carbon. Changing the carbon abundance will then affect the amount of available oxygen through the formation of more CO. However, in our test model we have $\log\epsilon_\mathrm O = 8.25$ and $\log\epsilon_\mathrm C = 7.42$ (see Table \ref{abundances1}), which means that oxygen is enhanced and carbon is reduced compared to the metallicity-scaled solar values. Thus, in our test model, which is supposed to be a model of a typical bulge giant, there is nearly seven times more oxygen than carbon, by number. Therefore, a change in the carbon abundance in a typical bulge star, with an enhanced oxygen abundance, will not change the strength of the OH lines as much as expected for a star with a solar-type carbon-to-oxygen ratio.

\item To summarise and to put it in another way, for our test model, the OH lines are mainly affected by the oxygen abundance, the CO lines by the C abundance, but
the CN lines are affected by the nitrogen, as well as the C and O abundances.


\end{itemize}

\begin{longtable}{llllccc}
\caption{\label{linelist} Line list of our 816 metal lines in the wavelength range observed. The seven columns  give (1) the wavelength in air, (2) the excitation energy of the lower level, (3) $\log gf$, (4) the radiation damping parameter (when no value  was 
available the very low value of $1.00\times10^5$ was used), (5) the van der Waals damping marked with an  `A'  when calculated according to \citet{anstee},  \citet{barklem:00} and references therein,  \citet{aspelund}, or  Barklem P., private communication (if a number is instead given this is an empirical correction factor to the van der Waals damping computed according to 
\citet{unsold}), (6) the atomic or ionic species, and (7) a star `*' when  the $\log gf$ value has been modified by spectrum fits to the 
   spectra of th Sun and Arcturus.}\\
\hline\hline
Wavelength [$\AA$] & $\chi_\mathrm{exc}$ [eV] & $\log gf$ & $\Gamma_\mathrm{rad}$ [rad\,s$^{-1}$] & van der Waal & Atomic/ionic species & New $\log gf$ values \\
\hline
\endfirsthead
\caption{continued.}\\
\hline\hline
Wavelength [$\AA$] & $\chi_\mathrm{exc}$ [eV] & $\log gf$ & $\Gamma_\mathrm{rad}$ [rad\,s$^{-1}$] & van der Waal & Atomic/ionic species & New $\log gf$ values \\
\hline
\endhead
\hline
\endfoot
 15315.660 & 6.280 & -1.170 & 1.00E+05    & A & Fe I \\  
 15320.084 & 4.770 & -2.307 & 4.69E+07    & A & Ti I \\  
 15320.897 & 7.030 & -2.080 & 1.00E+05    & A & Si I \\  
 15321.336 & 4.771 & -3.236 & 4.66E+07    & A & Ti I \\  
 15322.423 & 5.456 & -3.487 & 1.85E+08 & 2.50 & V  I \\  
 15322.705 & 6.554 & -1.037 & 8.11E+07 & 2.50 & Co I \\  
 15322.752 & 5.923 & -1.428 & 2.44E+08 & 2.50 & Cr I \\  
 15323.287 & 5.206 & -0.950 & 6.82E+07    & A & Ti I \\  
 15323.550 & 6.350 & -0.99  & 1.00E+05    & A & Fe I & * \\
 15323.809 & 6.730 & -3.652 & 4.09E+08    & A & Fe II  \\
 15323.973 & 6.412 & -3.406 & 4.12E+07    & A & Mn I \\  
 15324.020 & 6.127 & -2.719 & 8.79E+07    & A & Mn I \\  
 15325.641 & 4.650 & -3.349 & 1.07E+08 & 2.50 & V  I \\  
 15326.816 & 5.364 & -2.886 & 7.38E+07    & A & V  I \\  
 15327.159 & 5.221 & -3.407 & 6.19E+07    & A & Ti I \\  
 15327.168 & 6.412 & -3.018 & 4.34E+07    & A & Mn I \\  
 15327.518 & 7.125 & -2.980 & 1.00E+05    & A & Si I \\  
 15327.948 & 3.403 & -3.070 & 1.00E+05    & A & K  I \\  
 15328.033 & 2.845 & -6.324 & 1.96E+07    & A & Fe I \\  
 15329.048 & 4.851 & -0.673 & 1.25E+08    & A & V  I \\  
 15329.166 & 5.541 & -2.238 & 1.11E+08 & 2.50 & V  I \\  
 15329.358 & 3.403 & -3.370 & 1.00E+05    & A & K  I \\  
 15329.989 & 5.940 & -2.997 & 1.60E+08    & A & Co I \\  
 15330.017 & 4.396 & -3.084 & 1.10E+08    & A & Ti I \\  
 15330.191 & 6.718 & -1.90  & 1.00E+05    & A & Si I & * \\
 15331.188 & 4.865 & -0.265 & 1.39E+08 & 2.50 & V  I \\  
 15331.305 & 3.933 & -3.606 & 2.81E+08    & A & V  I \\  
 15331.775 & 4.585 & -1.942 & 6.40E+07 & 2.50 & V  I \\  
 15331.832 & 2.609 & -6.538 & 1.46E+07    & A & Fe I \\  
 15332.010 & 6.725 & -3.822 & 2.53E+08    & A & Fe I \\  
 15332.754 & 2.588 & -6.292 & 1.50E+07    & A & Fe I \\  
 15333.304 & 7.381 & -0.052 & 4.20E+08    & A & Sc II  \\
 15334.762 & 5.377 & -8.655 & 2.70E+08    & A & Mn II  \\
 15334.840 & 1.887 & -1.35  & 2.93E+06    & A & Ti I & * \\
 15335.011 & 5.607 & -2.997 & 1.50E+08    & A & Fe I \\  
 15335.380 & 5.410 &  0.00  & 1.12E+08    & A & Fe I & * \\
 15335.456 & 5.270 & -3.180 & 2.64E+08    & A & Ti I \\  
 15335.992 & 5.363 & -2.515 & 6.75E+08    & A & Ni I \\  
 15336.315 & 6.546 & -3.043 & 1.15E+08 & 2.50 & Co I \\  
 15336.338 & 4.865 & -1.200 & 1.39E+08    & A & V  I \\  
 15336.456 & 5.902 & -3.140 & 7.91E+07    & A & Fe I \\  
 15337.470 & 5.233 & -2.348 & 2.06E+08    & A & Ti I \\  
 15337.724 & 4.506 & -2.804 & 1.80E+08    & A & Ti I \\  
 15338.762 & 5.252 & -3.832 & 9.53E+07 & 2.50 & V  I \\  
 15338.780 & 6.261 & -2.580 & 1.00E+05    & A & Si I \\  
 15339.000 & 5.233 & -1.523 & 2.03E+08    & A & Ti I \\  
 15339.200 & 5.288 & -3.155 & 1.53E+08    & A & Ni I \\  
 15340.153 & 5.494 & -3.391 & 4.89E+08    & A & Ni I \\  
 15341.822 & 5.726 & -2.682 & 2.22E+08    & A & Co I \\  
 15342.140 & 4.256 & -1.299 & 1.85E+08    & A & Ti I \\  
 15342.199 & 5.851 & -3.552 & 2.25E+08    & A & Cr I \\  
 15342.973 & 7.108 & -1.85  & 1.00E+05    & A & Si I & * \\
 15343.070 & 9.330 & -3.900 & 1.00E+05    & A & C  I \\  
 15343.359 & 5.648 & -3.534 & 1.29E+08    & A & Fe I \\  
 15343.494 & 5.529 & -2.239 & 2.23E+08    & A & Sc I \\  
 15343.810 & 5.653 & -0.70  & 1.16E+08    & A & Fe I & * \\
 15344.883 & 4.586 & -1.470 & 6.52E+07 & 2.50 & V  I \\  
 15345.010 & 5.478 & -2.010 & 1.79E+08    & A & Fe I \\  
 15345.307 & 5.289 & -4.043 & 1.38E+08    & A & Ni I \\  
 15345.425 & 8.086 & -0.934 & 4.70E+08    & A & Ti II  \\
 15345.802 & 7.424 & -0.759 & 4.08E+08    & A & Sc II  \\
 15345.920 & 6.270 & -1.190 & 1.00E+05    & A & Fe I \\  
 15346.132 & 9.331 & -3.020 & 1.00E+05    & A & C  I \\  
 15346.827 & 4.850 & -3.648 & 1.85E+08    & A & Ti I \\  
 15348.252 & 5.464 & -1.785 & 8.22E+07 & 2.50 & V  I \\  
 15348.398 & 5.874 & -1.70  & 9.18E+07    & A & Fe I & * \\
 15348.723 & 4.835 & -3.379 & 2.06E+08    & A & Fe I \\  
 15348.904 & 3.397 & -1.070 & 1.00E+05    & A & K  I \\  
 15348.904 & 3.397 & -2.380 & 1.00E+05    & A & K  I \\  
 15348.950 & 5.950 & -1.00  & 1.00E+05    & A & Fe I & * \\
 15348.950 & 7.010 & -1.210 & 1.00E+05    & A & Mn I \\  
 15350.491 & 3.791 & -1.554 & 3.20E+08    & A & Sc I \\  
 15351.426 & 3.397 & -1.230 & 1.00E+05    & A & K  I \\  
 15352.410 & 5.458 & -3.227 & 1.20E+08    & A & Fe I \\  
 15352.612 & 5.059 & -2.307 & 7.03E+07    & A & Sc I \\  
 15352.919 & 5.377 & -7.823 & 2.70E+08    & A & Mn II  \\
 15353.155 & 8.098 & -1.983 & 4.74E+08    & A & Ti II  \\
 15354.098 & 4.598 & -2.165 & 1.43E+08    & A & Sc I \\  
 15354.756 & 6.223 & -4.020 & 1.00E+05    & A & Si I \\  
 15355.230 & 5.925 & -1.832 & 2.46E+08 & 2.50 & Cr I \\  
 15355.513 & 4.725 & -3.070 & 1.46E+08    & A & V  I \\  
 15355.560 & 4.712 & -2.525 & 1.39E+08    & A & V  I \\  
 15356.221 & 3.010 & -4.271 & 6.64E+06    & A & Cr I \\  
 15356.928 & 4.656 & -2.931 & 1.43E+08    & A & V  I \\  
 15357.046 & 4.652 & -3.345 & 2.61E+08    & A & V  I \\  
 15357.683 & 4.889 & -1.988 & 1.71E+08    & A & Mn I \\  
 15358.238 & 6.150 & -3.741 & 1.06E+08    & A & Fe I \\  
 15358.351 & 6.361 & -2.680 & 2.28E+08    & A & Fe I \\  
 15359.618 & 5.436 & -3.597 & 1.63E+08 & 2.50 & V  I \\  
 15359.722 & 5.228 & -7.048 & 5.79E+07    & A & Ca I \\  
 15359.736 & 3.799 & -5.272 & 2.84E+08    & A & V  II  \\
 15359.901 & 5.925 & -1.733 & 2.27E+08 & 2.50 & Cr I \\  
 15359.915 & 7.064 & -1.950 & 1.00E+05    & A & Si I \\  
 15360.019 & 6.127 & -2.457 & 8.79E+07    & A & Mn I \\  
 15360.230 & 4.260 & -2.97  & 8.75E+06    & A & Fe I & * \\
 15360.534 & 5.229 & -5.890 & 5.78E+07    & A & Ca I \\  
 15361.160 & 5.954 & -2.12  & 1.00E+05    & A & Si I & * \\
 15361.742 & 4.078 & -2.142 & 7.10E+07    & A & V  I \\  
 15362.049 & 4.888 & -3.677 & 1.08E+08    & A & Sc I \\  
 15363.045 & 5.026 & -4.761 & 1.99E+07    & A & Ca I \\  
 15363.890 & 6.022 & -3.993 & 4.71E+08 & 2.50 & Co I \\  
 15364.174 & 4.967 & -2.348 & 1.05E+08    & A & Sc I \\  
 15364.415 & 5.253 & -2.773 & 4.63E+07    & A & Ti I \\  
 15364.599 & 5.462 & -2.431 & 2.47E+08    & A & Cr I \\  
 15365.732 & 6.006 & -2.150 & 3.66E+08    & A & Co I \\  
 15365.874 & 5.330 & -7.100 & 2.63E+08    & A & Cr II  \\
 15365.968 & 5.812 & -3.902 & 2.64E+08    & A & Cr I \\  
 15366.521 & 5.228 & -6.343 & 5.52E+07    & A & Ca I \\  
 15366.677 & 5.936 & -3.223 & 7.35E+07 & 2.50 & Cr I \\  
 15367.334 & 5.229 & -4.789 & 5.51E+07    & A & Ca I \\  
 15367.560 & 5.228 & -4.827 & 5.52E+07    & A & Ca I \\  
 15367.740 & 4.922 & -0.545 & 9.10E+07    & A & Sc I \\  
 15368.104 & 5.228 & -4.111 & 5.52E+07    & A & Ca I \\  
 15368.373 & 5.229 & -3.916 & 5.51E+07    & A & Ca I \\  
 15368.439 & 5.229 & -3.712 & 5.51E+07    & A & Ca I \\  
 15369.684 & 5.814 & -3.556 & 1.44E+08    & A & Fe I \\  
 15370.953 & 8.089 & -0.255 & 4.83E+08    & A & Ti II  \\
 15371.320 & 5.870 & -2.240 & 1.00E+05    & A & Fe I \\  
 15371.430 & 5.303 & -3.310 & 1.29E+08    & A & Ni I \\  
 15371.473 & 4.498 & -1.122 & 6.10E+07    & A & Sc I \\  
 15371.590 & 5.303 & -2.430 & 1.29E+08    & A & Ni I \\  
 15371.750 & 5.303 & -2.010 & 1.29E+08    & A & Ni I \\  
 15372.702 & 4.059 & -2.259 & 2.72E+08    & A & V  I \\  
 15372.986 & 6.006 & -1.829 & 3.24E+08    & A & Co I \\  
 15374.256 & 5.125 & -2.571 & 1.19E+08    & A & Ti I \\  
 15374.263 & 6.212 & -2.826 & 3.57E+08 & 2.50 & Mn I \\  
 15375.340 & 5.920 & -1.870 & 1.00E+05    & A & Fe I \\  
 15375.412 & 5.026 & -2.856 & 2.01E+07    & A & Ca I \\  
 15375.430 & 6.734 & -1.530 & 1.00E+05    & A & Si I \\  
 15375.440 & 5.202 & -3.635 & 1.50E+08    & A & Ti I \\  
 15376.249 & 5.948 & -1.826 & 1.55E+08 & 2.50 & Cr I \\  
 15376.348 & 5.502 & -3.360 & 1.54E+08    & A & Fe I \\  
 15376.500 & 4.733 & -4.085 & 2.07E+08    & A & Fe I \\  
 15376.830 & 6.223 & -0.61  & 1.00E+05    & A & Si I & * \\
 15376.830 & 6.721 & -1.08  & 1.00E+05    & A & Si I & * \\
 15376.888 & 4.579 & -2.950 & 1.86E+08    & A & Sc I \\  
 15378.614 & 3.011 & -3.962 & 6.64E+06    & A & Cr I \\  
 15378.676 & 5.280 & -2.524 & 1.34E+08    & A & Ni I \\  
 15379.040 & 6.280 & -3.236 & 1.91E+08    & A & Fe I \\  
 15379.296 & 5.446 & -3.287 & 2.29E+08    & A & Fe I \\  
 15379.513 & 4.929 & -1.175 & 1.22E+08    & A & Sc I \\  
 15379.750 & 8.116 & -1.403 & 4.78E+08    & A & Ti II  \\
 15380.117 & 5.481 & -2.113 & 2.13E+08    & A & Ni I \\  
 15381.044 & 5.203 & -2.856 & 3.35E+08    & A & Ti I \\  
 15381.080 & 5.230 & -2.645 & 8.02E+07    & A & Ti I \\  
 15381.110 & 2.333 & -2.29  & 7.13E+07    & A & Ti I & * \\
 15381.217 & 3.640 & -6.217 & 9.86E+07    & A & Fe I \\  
 15381.288 & 8.114 & -0.396 & 5.08E+08    & A & Ti II  \\
 15381.738 & 6.721 & -2.030 & 1.00E+05    & A & Si I \\  
 15381.980 & 3.640 & -3.03  & 1.00E+05    & A & Fe I & * \\
 15382.753 & 2.305 & -6.848 & 1.73E+06    & A & Ti I \\  
 15382.992 & 7.951 & -2.741 & 6.10E+08    & A & V  II  \\
 15384.110 & 6.200 & -1.460 & 1.00E+05    & A & Fe I \\  
 15387.069 & 7.166 & -1.640 & 1.00E+05    & A & Si I \\  
 15387.635 & 4.769 & -5.281 & 1.58E+08    & A & Ca I \\  
 15387.800 & 6.280 & -0.270 & 1.00E+05    & A & Fe I \\  
 15388.746 & 4.650 & -2.411 & 1.51E+09    & A & V  I \\  
 15388.976 & 5.140 & -3.602 & 1.15E+08    & A & Ti I \\  
 15389.819 & 5.219 & -0.859 & 7.11E+07    & A & Ti I \\  
 15390.028 & 5.497 & -3.477 & 3.29E+08    & A & Ni I \\  
 15390.428 & 4.579 & -1.511 & 1.85E+08    & A & Sc I \\  
 15390.663 & 5.853 & -1.302 & 3.54E+08    & A & Ni I \\  
 15390.826 & 4.669 & -0.777 & 1.19E+08    & A & Ti I \\  
 15390.902 & 8.599 & -2.858 & 6.17E+08    & A & V  II  \\
 15391.281 & 4.975 & -1.837 & 7.48E+07 & 2.50 & Sc I \\  
 15391.968 & 5.161 & -2.020 & 1.18E+08    & A & Ti I \\  
 15394.670 & 5.621 & -0.03  & 1.61E+08    & A & Fe I & * \\
 15394.680 & 2.469 & -9.023 & 5.82E+03    & A & Fe I \\  
 15394.777 & 5.045 & -1.729 & 4.20E+07    & A & Ca I \\  
 15395.720 & 5.621 & -0.23  & 1.21E+08    & A & Fe I & * \\
 15396.093 & 5.122 & -3.938 & 3.70E+08    & A & Co II  \\
 15397.357 & 4.769 & -5.386 & 1.58E+08    & A & Ca I \\  
 15397.629 & 4.398 & -2.926 & 1.19E+07    & A & Ti I \\  
 15397.753 & 4.885 & -1.264 & 1.36E+08    & A & V  I \\  
 15399.285 & 2.334 & -2.130 & 7.13E+07    & A & Ti I \\  
 15399.605 & 5.621 & -3.216 & 1.21E+08    & A & Fe I \\  
 15399.989 & 4.515 & -2.449 & 1.83E+08    & A & Ti I \\  
 15400.060 & 8.700 &  0.100 & 1.00E+05    & A & S  I \\  
 15400.082 & 8.307 & -3.765 & 4.98E+08    & A & Fe II  \\
 15401.833 & 6.728 & -3.525 & 2.53E+08    & A & Fe I \\  
 15402.065 & 4.769 & -5.265 & 1.58E+08    & A & Ca I \\  
 15402.331 & 7.439 & -2.555 & 3.99E+08    & A & Sc II  \\
 15402.331 & 8.699 & -1.60  & 1.00E+05    & A & S  I & * \\
 15403.072 & 4.770 & -2.457 & 7.53E+07    & A & Ti I \\  
 15403.494 & 8.071 & -0.300 & 4.89E+08    & A & Ti II  \\
 15403.770 & 8.700 &  0.35  & 1.00E+05    & A & S  I & * \\
 15403.921 & 5.011 & -9.415 & 2.50E+08    & A & Cr II  \\
 15404.206 & 4.638 & -3.270 & 2.62E+08    & A & V  I \\  
 15405.191 & 5.214 & -2.857 & 1.35E+08    & A & Ti I \\  
 15405.227 & 3.823 & -4.374 & 4.06E+07    & A & V  I \\  
 15405.227 & 4.664 & -1.567 & 1.50E+08    & A & V  I \\  
 15405.979 & 8.700 & -1.50  & 1.00E+05    & A & S  I & * \\
 15406.485 & 2.950 & -1.121 & 1.00E+05    & A & Rb I \\  
 15406.699 & 5.700 & -3.446 & 3.09E+08    & A & Co I \\  
 15406.865 & 5.464 & -0.996 & 2.47E+08    & A & Cr I \\  
 15407.048 & 5.967 & -3.607 & 1.63E+08    & A & Fe I \\  
 15407.079 & 5.062 & -1.002 & 6.98E+07    & A & Sc I \\  
 15407.143 & 5.614 & -3.761 & 1.40E+08    & A & Fe I \\  
 15407.435 & 5.521 & -2.728 & 1.02E+08 & 2.50 & V  I \\  
 15408.560 & 5.648 & -2.290 & 1.51E+08    & A & Fe I \\  
 15408.864 & 5.283 & -2.953 & 1.52E+08    & A & Ti I \\  
 15409.335 & 8.583 & -3.623 & 6.41E+08    & A & V  II  \\
 15410.071 & 2.950 & -2.075 & 1.00E+05    & A & Rb I \\  
 15410.261 & 3.013 & -4.442 & 6.64E+06    & A & Cr I \\  
 15411.513 & 5.250 & -6.097 & 1.13E+08    & A & Ca I \\  
 15411.627 & 5.519 & -3.159 & 1.69E+08    & A & Fe I \\  
 15412.254 & 6.726 & -2.60  & 1.00E+05    & A & Mg I & * \\
 15412.748 & 5.250 & -3.896 & 1.13E+08    & A & Ca I \\  
 15415.289 & 6.726 & -1.870 & 1.00E+05    & A & Mg I \\  
 15415.289 & 6.726 & -2.15  & 1.00E+05    & A & Mg I & * \\
 15415.346 & 5.445 & -1.693 & 4.42E+08 & 2.50 & V  I \\  
 15415.850 & 5.980 & -2.100 & 1.00E+05    & A & Fe I \\  
 15415.917 & 4.650 & -0.189 & 8.99E+07 & 2.50 & V  I \\  
 15415.964 & 4.925 & -1.971 & 6.52E+07    & A & Sc I \\  
 15416.428 & 5.538 & -2.999 & 2.74E+08    & A & Fe I \\  
 15416.485 & 5.538 & -3.207 & 2.89E+08    & A & Fe I \\  
 15416.589 & 4.771 & -2.058 & 7.53E+07    & A & Ti I \\  
 15416.965 & 3.642 & -7.769 & 9.86E+07    & A & Fe I \\  
 15417.231 & 4.424 & -1.232 & 1.02E+08    & A & Ti I \\  
 15417.578 & 5.607 & -4.078 & 1.32E+08    & A & Fe I \\  
 15418.294 & 5.541 & -2.888 & 1.01E+08 & 2.50 & V  I \\  
 15418.461 & 4.706 & -2.005 & 2.55E+08    & A & V  I \\  
 15418.865 & 5.855 & -3.478 & 2.43E+08    & A & Cr I \\  
 15419.079 & 1.804 & -3.958 & 1.09E+07    & A & V  I \\  
 15419.643 & 5.250 & -0.779 & 1.26E+08    & A & Ca I \\  
 15420.649 & 5.089 & -2.600 & 6.41E+07    & A & Co I \\  
 15420.806 & 7.006 & -2.33 & 1.00E+05 & 1.30 & Si I & * \\
 15420.884 & 5.502 & -3.442 & 2.45E+08    & A & Fe I \\  
 15421.567 & 6.727 & -1.70  & 1.00E+05    & A & Mg I & * \\
 15421.567 & 6.727 & -1.790 & 1.00E+05    & A & Mg I \\  
 15421.567 & 6.727 & -3.340 & 1.00E+05    & A & Mg I \\  
 15421.662 & 7.006 & -2.660 & 1.00E+05 & 1.30 & Si I \\  
 15422.076 & 8.056 & -0.554 & 4.95E+08    & A & Ti II  \\
 15422.260 & 8.701 &  0.55  & 1.00E+05    & A & S  I & * \\
 15422.260 & 8.701 & -0.620 & 1.00E+05    & A & S  I \\  
 15422.680 & 6.340 & -0.910 & 1.00E+05    & A & Fe I \\  
 15422.980 & 5.599 & -3.876 & 1.73E+08    & A & Mn I \\  
 15424.455 & 7.772 & -4.602 & 2.88E+08    & A & Cr II  \\
 15424.455 & 8.701 & -1.920 & 1.00E+05    & A & S  I \\  
 15425.348 & 5.188 & -2.554 & 2.64E+08    & A & Ti I \\  
 15426.520 & 6.160 & -2.860 & 1.00E+05    & A & Fe I \\  
 15426.970 & 1.873 & -2.52  & 2.76E+06    & A & Ti I & * \\
 15427.097 & 5.782 & -2.530 & 1.23E+08    & A & Cr I \\  
 15427.407 & 7.757 & -5.217 & 3.08E+08    & A & Mn II  \\
 15427.610 & 6.450 & -0.980 & 1.00E+05    & A & Fe I \\  
 15428.478 & 4.597 & -1.034 & 8.17E+07 & 2.50 & V  I \\  
 15430.050 & 5.140 & -2.556 & 1.19E+08    & A & Ti I \\  
 15431.450 & 7.134 & -3.590 & 1.00E+05    & A & Si I \\  
 15431.650 & 6.261 & -3.720 & 1.00E+05    & A & Si I \\  
 15431.955 & 6.201 & -0.771 & 2.62E+08    & A & Mn I \\  
 15432.453 & 4.854 & -2.753 & 1.85E+08    & A & Ti I \\  
 15434.409 & 5.856 & -2.675 & 2.31E+08    & A & Cr I \\  
 15434.433 & 5.570 & -3.485 & 9.82E+07 & 2.50 & V  I \\  
 15434.981 & 5.855 & -3.045 & 2.31E+08    & A & Co I \\  
 15435.360 & 7.125 & -2.240 & 1.00E+05    & A & Si I \\  
 15436.261 & 5.900 & -2.611 & 1.27E+08    & A & Fe I \\  
 15437.320 & 5.840 & -1.910 & 1.00E+05    & A & Fe I \\  
 15437.467 & 4.607 & -4.030 & 2.62E+08    & A & Fe I \\  
 15437.477 & 5.839 & -2.552 & 1.29E+08    & A & Fe I \\  
 15438.135 & 2.740 & -5.022 & 3.00E+05    & A & Ni I \\  
 15438.194 & 4.411 & -1.972 & 2.17E+08    & A & Ti I \\  
 15438.480 & 1.502 & -4.626 & 1.69E+06    & A & Ti I \\  
 15440.047 & 2.305 & -7.337 & 1.72E+06    & A & Ti I \\  
 15440.354 & 5.943 & -3.227 & 1.35E+08    & A & Fe I \\  
 15441.177 & 6.552 & -1.510 & 1.00E+05    & A & Cu I \\  
 15441.418 & 4.777 & -3.611 & 3.92E+07    & A & Ti I \\  
 15441.800 & 5.874 & -1.840 & 1.09E+08    & A & Fe I \\  
 15441.957 & 3.415 & -5.436 & 1.65E+08    & A & Fe I \\  
 15442.682 & 5.526 & -3.960 & 3.89E+08    & A & Co I \\  
 15443.517 & 4.063 & -4.145 & 6.68E+07    & A & V  I \\  
 15444.185 & 5.464 & -1.353 & 5.02E+08 & 2.50 & V  I \\  
 15445.283 & 7.758 & -4.023 & 3.08E+08    & A & Mn II  \\
 15445.381 & 5.614 & -3.546 & 1.38E+08    & A & Fe I \\  
 15445.855 & 4.978 & -2.553 & 1.02E+08 & 2.50 & Sc I \\  
 15447.526 & 4.603 & -2.074 & 1.44E+08    & A & Sc I \\  
 15447.836 & 4.627 & -3.415 & 2.57E+08    & A & V  I \\  
 15448.352 & 5.481 & -3.263 & 2.15E+08    & A & Ni I \\  
 15450.869 & 6.125 & -1.758 & 1.53E+08    & A & Ni I \\  
 15451.330 & 6.450 & -0.48  & 1.00E+05    & A & Fe I & * \\
 15451.940 & 6.290 & -1.290 & 1.00E+05    & A & Fe I \\  
 15451.940 & 6.340 & -1.290 & 1.00E+05    & A & Fe I \\  
 15452.755 & 5.234 & -2.386 & 1.04E+08    & A & Ti I \\  
 15453.080 & 7.069 & -3.120 & 1.00E+05    & A & Si I \\  
 15453.343 & 3.867 & -1.915 & 6.49E+07    & A & Ti I \\  
 15454.045 & 5.036 & -4.005 & 1.21E+08    & A & Co I \\  
 15454.356 & 6.212 & -0.772 & 2.51E+08    & A & Mn I \\  
 15454.456 & 5.283 & -2.732 & 1.36E+08    & A & Ti I \\  
 15456.028 & 6.185 & -1.124 & 2.86E+08    & A & Mn I \\  
 15456.270 & 6.340 & -1.790 & 1.00E+05    & A & Fe I \\  
 15456.666 & 2.450 & -9.207 & 7.83E+03    & A & Fe I \\  
 15456.728 & 5.648 & -3.730 & 1.30E+08    & A & Fe I \\  
 15456.733 & 5.913 & -2.697 & 1.29E+08    & A & Fe I \\  
 15459.398 & 4.049 & -3.275 & 7.10E+07    & A & V  I \\  
 15461.191 & 5.510 & -3.356 & 1.26E+08 & 2.50 & V  I \\  
 15461.220 & 4.186 & -4.341 & 3.97E+08    & A & Fe I \\  
 15462.267 & 5.978 & -2.418 & 3.52E+08    & A & Co I \\  
 15462.420 & 6.290 & -2.070 & 1.00E+05    & A & Fe I \\  
 15462.650 & 5.941 & -2.394 & 9.33E+07 & 2.50 & Cr I \\  
 15463.033 & 5.537 & -4.067 & 3.48E+08    & A & V  II  \\
 15463.710 & 5.284 & -1.640 & 4.29E+08    & A & Ni I \\  
 15464.820 & 5.049 & -1.620 & 7.96E+07    & A & Ca I \\  
 15466.023 & 6.412 & -2.174 & 6.37E+07    & A & Mn I \\  
 15466.645 & 9.007 & -1.869 & 2.89E+08    & A & Mn II  \\
 15466.813 & 5.464 & -1.141 & 9.59E+07 & 2.50 & V  I \\  
 15467.729 & 4.640 & -0.164 & 8.83E+07    & A & V  I \\  
 15467.985 & 6.028 & -3.920 & 4.02E+08 & 2.50 & Co I \\  
 15468.122 & 4.396 & -3.628 & 2.00E+08    & A & Ti I \\  
 15468.655 & 6.212 & -2.323 & 3.44E+08 & 2.50 & Mn I \\  
 15469.820 & 8.045 & -0.45  & 1.00E+05    & A & S  I & * \\
 15470.293 & 6.803 & -4.600 & 3.07E+08    & A & Fe II  \\
 15470.570 & 4.394 & -3.525 & 2.87E+08    & A & Ti I \\  
 15471.121 & 5.953 & -1.627 & 2.55E+08 & 2.50 & Cr I \\  
 15471.964 & 6.726 & -2.40  & 1.00E+05    & A & Si I & * \\
 15472.098 & 4.355 & -3.052 & 2.21E+08    & A & Ti I \\  
 15472.701 & 6.006 & -3.184 & 3.14E+08    & A & Co I \\  
 15473.252 & 4.644 & -3.500 & 1.02E+08    & A & V  I \\  
 15473.324 & 5.760 & -3.340 & 2.01E+08    & A & Co I \\  
 15474.603 & 5.902 & -3.659 & 1.39E+08    & A & Fe I \\  
 15474.641 & 5.480 & -3.149 & 3.50E+08    & A & V  II  \\
 15474.977 & 7.430 & -1.194 & 4.05E+08    & A & Sc II  \\
 15475.190 & 5.486 & -2.110 & 1.36E+08    & A & Fe I \\  
 15475.190 & 6.310 & -0.730 & 1.00E+05    & A & Fe I \\  
 15475.620 & 8.046 & -0.890 & 1.00E+05    & A & S  I \\  
 15475.808 & 3.967 & -9.036 & 2.94E+08    & A & Fe II  \\
 15475.887 & 4.585 & -1.706 & 7.29E+07 & 2.50 & V  I \\  
 15475.923 & 5.874 & -2.00  & 1.26E+08    & A & Fe I & * \\
 15476.500 & 6.320 & -1.140 & 1.00E+05    & A & Fe I \\  
 15476.917 & 4.690 & -0.055 & 1.39E+08    & A & Ti I \\  
 15478.480 & 8.046 & -0.220 & 1.00E+05    & A & S  I \\  
 15478.870 & 6.240 & -0.600 & 1.00E+05    & A & Fe I \\  
 15478.978 & 4.741 & -3.181 & 1.32E+08    & A & V  I \\  
 15479.050 & 6.101 & -3.075 & 2.66E+08    & A & V  II  \\
 15479.600 & 6.320 & -1.120 & 1.00E+05    & A & Fe I \\  
 15480.081 & 4.232 & -3.345 & 1.02E+08    & A & V  I \\  
 15480.230 & 5.614 & -2.270 & 1.56E+08    & A & Fe I \\  
 15480.967 & 2.328 & -8.788 & 1.96E+04    & A & Co I \\  
 15481.087 & 7.439 & -1.140 & 4.02E+08    & A & Sc II  \\
 15481.878 & 7.760 & -3.150 & 3.08E+08    & A & Mn II  \\
 15481.936 & 4.433 & -1.610 & 2.00E+08    & A & Ti I \\  
 15484.420 & 4.650 & -0.948 & 9.04E+07 & 2.50 & V  I \\  
 15484.495 & 6.582 & -2.973 & 5.66E+07    & A & Fe I \\  
 15485.450 & 6.280 & -0.93  & 1.00E+05    & A & Fe I & * \\
 15485.452 & 5.140 & -3.360 & 1.19E+08    & A & Ti I \\  
 15486.627 & 4.168 & -2.265 & 2.04E+08    & A & Sc I \\  
 15487.097 & 5.241 & -3.031 & 7.69E+07    & A & Ti I \\  
 15488.590 & 5.928 & -3.431 & 1.04E+08    & A & Fe I \\  
 15488.657 & 5.241 & -3.578 & 7.40E+07    & A & Ti I \\  
 15488.813 & 7.104 & -2.50  & 1.00E+05    & A & Si I & * \\
 15489.425 & 5.181 & -2.749 & 5.61E+07    & A & Ca I \\  
 15489.458 & 6.421 & -1.935 & 4.12E+07    & A & Mn I \\  
 15489.626 & 4.967 & -1.708 & 1.06E+08    & A & Sc I \\  
 15489.658 & 4.420 & -2.739 & 1.54E+08    & A & Ti I \\  
 15489.674 & 6.054 & -3.104 & 1.66E+08    & A & Co I \\  
 15489.960 & 5.879 & -3.509 & 1.77E+08    & A & Fe I \\  
 15489.986 & 8.093 & -1.216 & 4.69E+08    & A & Ti II  \\
 15490.073 & 3.869 & -2.881 & 6.67E+07    & A & Ti I \\  
 15490.340 & 2.198 & -4.85 & 1.14E+04 & 1.40 & Fe I & * \\
 15490.490 & 5.319 & -6.527 & 2.35E+08    & A & Cr II  \\
 15490.880 & 6.290 & -0.57  & 1.00E+05    & A & Fe I & * \\
 15490.971 & 7.832 & -1.551 & 1.21E+09    & A & Ti II  \\
 15492.140 & 5.839 & -1.950 & 1.27E+08    & A & Fe I \\  
 15493.515 & 6.368 & -1.45  & 2.34E+08    & A & Fe I & * \\
 15493.550 & 6.450 & -1.25  & 1.00E+05    & A & Fe I & * \\
 15494.800 & 5.177 & -3.163 & 3.30E+08    & A & Ti I \\  
 15495.244 & 7.472 & -2.980 & 4.50E+08    & A & Sc II  \\\
 15495.364 & 4.929 & -2.838 & 1.18E+08    & A & Sc I \\  
 15495.456 & 5.168 & -2.201 & 9.59E+07    & A & Ca I \\  
 15495.667& 7.006 & -3.110 & 1.00E+05  & 1.30 & Si I \\  
 15496.690 & 6.290 & -0.300 & 1.00E+05    & A & Fe I \\  
 15496.830 & 5.102 & -2.286 & 7.74E+07 & 2.50 & Sc I \\  
 15496.964 & 7.006 & -2.54 & 1.00E+05 & 1.30 & Si I & * \\
 15497.000 & 6.269 & -3.400 & 1.00E+05    & A & Si I \\  
 15497.040 & 6.290 & -1.140 & 1.00E+05    & A & Fe I \\  
 15499.232 & 5.506 & -3.100 & 2.30E+08    & A & Fe I \\  
 15499.410 & 6.350 & -0.32  & 1.00E+05    & A & Fe I & * \\
 15499.689 & 5.796 & -2.975 & 2.25E+08    & A & Co I \\  
 15500.073 & 4.504 & -0.244 & 5.75E+07    & A & Sc I \\  
 15500.241 & 9.036 & -3.901 & 8.30E+08    & A & V  II  \\
 15500.316 & 4.361 & -2.823 & 8.11E+07    & A & Ti I \\  
 15500.650 & 6.201 & -1.332 & 3.12E+08 & 2.50 & Mn I \\  
 15500.800 & 6.320 & -0.12  & 1.00E+05    & A & Fe I & * \\
 15501.080 & 5.943 & -3.981 & 1.92E+08    & A & Fe I \\  
 15501.320 & 6.290 &  0.10  & 1.00E+05    & A & Fe I & * \\
 15502.170 & 6.350 & -1.070 & 1.00E+05    & A & Fe I \\  
 15502.429 & 3.409 & -4.954 & 1.46E+08    & A & Co I \\  
 15502.640 & 7.134 & -1.870 & 1.00E+05    & A & Si I \\  
 15503.246 & 3.375 & -4.682 & 5.86E+06    & A & Cr I \\  
 15503.246 & 4.168 & -1.810 & 2.14E+08    & A & Sc I \\  
 15503.840 & 5.972 & -3.183 & 3.33E+08    & A & Fe I \\  
 15503.943 & 5.726 & -3.178 & 2.24E+08    & A & Co I \\  
 15503.967 & 4.725 & -2.564 & 1.05E+08    & A & V  I \\  
 15505.771 & 4.389 & -2.911 & 1.22E+07    & A & Ti I \\  
 15506.105 & 5.524 & -3.367 & 1.67E+08    & A & Fe I \\  
 15506.252 & 4.968 & -2.486 & 9.68E+07    & A & Sc I \\  
 15506.980 & 6.727 & -1.80  & 1.00E+05    & A & Si I & * \\
 15507.022 & 4.865 & -0.479 & 1.44E+08    & A & V  I \\  
 15507.043 & 8.228 & -1.500 & 1.00E+05    & A & P  I \\  
 15507.103 & 4.770 & -2.115 & 4.91E+07    & A & Ti I \\  
 15507.118 & 8.943 & -3.497 & 3.84E+08    & A & Fe II  \\
 15507.623 & 5.235 & -0.895 & 6.70E+07    & A & Ti I \\  
 15508.385 & 4.771 & -3.338 & 4.89E+07    & A & Ti I \\  
 15511.117 & 7.166 & -2.850 & 1.00E+05    & A & Si I \\  
 15511.528 & 5.476 & -3.146 & 2.28E+08    & A & Fe I \\  
 15512.724 & 4.681 & -1.408 & 8.93E+07 & 2.50 & V  I \\  
 15513.511 & 4.515 & -2.559 & 1.81E+08    & A & Ti I \\  
 15514.280 & 6.290 & -0.750 & 1.00E+05    & A & Fe I \\  
 15514.691 & 7.092 & -2.270 & 1.00E+05    & A & Si I \\  
 15515.368 & 2.305 & -4.104 & 1.72E+06    & A & Ti I \\  
 15515.373 & 4.646 & -2.758 & 1.10E+08    & A & V  I \\  
 15515.760 & 6.290 & -1.700 & 1.00E+05    & A & Fe I \\  
 15515.876 & 4.793 & -2.070 & 2.27E+08    & A & Ti I \\  
 15516.720 & 6.290 & -1.330 & 1.00E+05    & A & Fe I \\  
 15517.275 & 5.744 & -3.579 & 3.07E+08    & A & Co I \\  
 15518.720 & 5.104 & -0.756 & 7.76E+07 & 2.50 & Sc I \\  
 15518.900 & 6.280 & -1.380 & 1.00E+05    & A & Fe I \\  
 15519.100 & 6.290 & -1.160 & 1.00E+05    & A & Fe I \\  
 15519.360 & 6.290 & -0.570 & 1.00E+05    & A & Fe I \\  
 15519.636 & 4.111 & -1.236 & 3.13E+08    & A & V  I \\  
 15519.942 & 2.484 & -5.827 & 1.40E+07    & A & Fe I \\  
 15519.949 & 3.807 & -1.148 & 3.16E+08    & A & Sc I \\  
 15520.115 & 7.108 & -1.85  & 1.00E+05    & A & Si I & * \\
 15521.086 & 5.352 & -3.473 & 1.98E+08    & A & Fe I \\  
 15521.690 & 6.320 & -1.440 & 1.00E+05    & A & Fe I \\  
 15522.600 & 6.212 & -2.760 & 1.07E+08    & A & Co I \\  
 15522.640 & 6.320 & -0.97  & 1.00E+05    & A & Fe I & * \\
 15523.998 & 6.054 & -3.914 & 1.78E+08    & A & Co I \\  
 15524.277 & 2.740 & -7.135 & 4.63E+04    & A & Ni I \\  
 15524.300 & 5.790 & -1.510 & 1.00E+05    & A & Fe I \\  
 15524.543 & 5.793 & -2.15  & 2.55E+08    & A & Fe I & * \\
 15525.227 & 5.840 & -2.738 & 1.17E+08    & A & Fe I \\  
 15525.661 & 4.885 & -0.054 & 1.63E+08 & 2.50 & V  I \\  
 15525.734 & 8.097 & -0.400 & 5.12E+08    & A & Ti II  \\
 15525.738 & 4.793 & -1.559 & 2.27E+08    & A & Ti I \\  
 15527.210 & 6.320 & -1.010 & 1.00E+05    & A & Fe I \\  
 15527.535 & 7.140 & -3.220 & 1.00E+05    & A & Si I \\  
 15528.109 & 5.947 & -2.974 & 4.53E+08    & A & Fe I \\  
 15530.043 & 3.573 & -6.583 & 9.31E+07    & A & Fe I \\  
 15530.195 & 5.445 & -1.530 & 8.15E+07 & 2.50 & V  I \\  
 15531.288 & 4.654 & -0.201 & 1.93E+08    & A & Ti I \\  
 15531.750 & 5.642 & -0.48  & 1.22E+08    & A & Fe I & * \\
 15532.263 & 7.140 & -2.450 & 1.00E+05    & A & Si I \\  
 15532.449 & 6.718 & -2.18  & 1.00E+05    & A & Si I & * \\
 15533.977 & 7.140 & -3.660 & 1.00E+05    & A & Si I \\  
 15534.260 & 5.642 & -0.30  & 1.21E+08    & A & Fe I & * \\
 15537.450 & 5.790 & -1.710 & 1.00E+05    & A & Fe I \\  
 15537.572 & 5.793 & -1.410 & 2.56E+08    & A & Fe I \\  
 15537.690 & 6.320 & -0.50  & 1.71E+08    & A & Fe I & * \\
 15538.064 & 5.744 & -1.809 & 3.21E+08    & A & Co I \\  
 15538.463 & 6.761 & -2.36  & 1.00E+05    & A & Si I & * \\
 15539.417 & 5.957 & -4.069 & 6.44E+07 & 2.50 & Cr I \\  
 15541.547 & 5.844 & -3.014 & 2.15E+08    & A & Fe I \\  
 15541.852 & 5.967 & -2.837 & 2.25E+08    & A & Fe I \\  
 15541.857 & 6.371 & -1.868 & 1.93E+08    & A & Fe I \\  
 15542.016 & 7.006 & -1.380 & 1.00E+05 & 1.30&  Si I \\  
 15542.090 & 5.642 & -0.700 & 1.64E+08    & A & Fe I \\  
 15542.090 & 7.010 & -1.380 & 1.00E+05    & A & Si I \\  
 15542.197 & 4.690 & -0.726 & 1.17E+08    & A & Ti I \\  
 15542.731 & 4.394 & -3.457 & 2.79E+08    & A & Ti I \\  
 15543.357 & 4.860 & -1.938 & 1.85E+08    & A & Ti I \\  
 15543.780 & 1.879 & -1.45  & 2.76E+06    & A & Ti I & * \\
 15543.838 & 4.795 & -1.310 & 2.40E+08    & A & Ti I \\  
 15544.152 & 6.792 & -1.440 & 1.00E+05    & A & Cu I \\  
 15544.355 & 2.487 & -7.407 & 7.66E+07    & A & Ti I \\  
 15546.081 & 4.405 & -1.930 & 2.26E+08    & A & Ti I \\  
 15548.978 & 5.181 & -6.986 & 5.38E+07    & A & Ca I \\  
 15550.450 & 6.340 & -0.340 & 1.71E+08    & A & Fe I \\  
 15550.560 & 6.112 & -3.186 & 6.25E+07    & A & Fe I \\  
 15551.430 & 6.350 & -0.290 & 1.71E+08    & A & Fe I \\  
 15551.818 & 4.664 & -2.093 & 1.65E+08    & A & V  I \\  
 15552.108 & 8.108 & -1.104 & 4.71E+08    & A & Ti II  \\
 15552.222 & 5.621 & -4.20  & 1.09E+08    & A & Fe I & * \\
 15553.245 & 5.537 & -4.569 & 2.27E+08    & A & V  II  \\
 15553.560 & 5.478 & -3.056 & 2.65E+08    & A & Fe I \\  
 15554.510 & 6.280 & -1.20  & 1.00E+05    & A & Fe I & * \\
 15554.510 & 6.410 & -1.240 & 1.00E+05    & A & Fe I \\  
 15554.557 & 4.433 & -1.828 & 1.76E+08    & A & Ti I \\  
 15554.625 & 4.607 & -3.570 & 9.79E+07 & 2.50 & V  I \\  
 15554.697 & 5.868 & -3.799 & 3.67E+08    & A & V  II  \\
 15555.120 & 5.280 & -0.610 & 1.00E+05    & A & Ni I \\  
 15555.210 & 5.280 & -1.030 & 1.00E+05    & A & Ni I \\  
 15555.370 & 5.488 &  0.08  & 1.48E+08    & A & Ni I & * \\
 15555.641 & 5.059 & -1.139 & 7.35E+07    & A & Sc I \\  
 15556.016 & 5.283 & -3.15  & 1.68E+08    & A & Ni I & * \\
 15556.122 & 6.552 & -1.320 & 1.00E+05    & A & Cu I \\  
 15556.670 & 5.930 & -2.804 & 1.38E+08    & A & Fe I \\  
 15556.711 & 5.297 & -2.788 & 1.55E+08    & A & Ti I \\  
 15557.602 & 4.681 & -4.320 & 1.80E+08    & A & V  I \\  
 15557.689 & 5.289 & -2.010 & 8.97E+07    & A & Ti I \\  
 15557.790 & 5.964 & -0.62  & 1.00E+05    & A & Si I & * \\
 15558.585 & 4.500 & -2.580 & 2.05E+08    & A & Ti I \\  
 15559.500 & 5.868 & -2.459 & 1.24E+08    & A & Ni I \\  
 15559.849 & 5.928 & -3.843 & 1.04E+08    & A & Fe I \\  
 15559.922 & 5.168 & -6.944 & 9.06E+07    & A & Ca I \\  
 15560.780 & 6.350 & -0.510 & 1.00E+05    & A & Fe I \\  
 15561.041 & 6.078 & -4.003 & 2.57E+08    & A & Co I \\  
 15561.251 & 7.040 & -1.380 & 1.00E+05    & A & Si I \\  
 15561.268 & 6.711 & -3.630 & 4.45E+08    & A & Fe I \\  
 15562.080 & 4.627 & -4.159 & 5.87E+07    & A & V  I \\  
 15562.300 & 6.366 & -2.677 & 1.81E+08    & A & Ni I \\  
 15563.463 & 5.237 & -3.380 & 5.13E+07    & A & Cr I \\  
 15564.020 & 9.053 & -3.801 & 8.38E+08    & A & Fe II  \\
 15564.369 & 5.614 & -2.378 & 1.38E+08    & A & Fe I \\  
 15564.723 & 4.527 & -2.044 & 3.28E+07    & A & Sc I \\  
 15565.230 & 6.320 & -0.950 & 1.00E+05    & A & Fe I \\  
 15565.886 & 4.640 & -0.903 & 9.82E+07 & 2.50 & V  I \\  
 15566.725 & 6.350 & -0.53  & 1.71E+08    & A & Fe I & * \\
 15567.001 & 4.967 & -1.920 & 1.07E+08    & A & Sc I \\  
 15567.188 & 7.113  & -3.830 & 1.00E+05 & 1.30 & Si I \\  
 15567.261 & 6.350 & -3.891 & 1.71E+08    & A & Fe I \\  
 15567.680 & 4.623 & -3.966 & 2.27E+08    & A & V  I \\  
 15567.728 & 4.586 & -3.805 & 1.10E+08 & 2.50 & V  I \\  
 15568.187 & 7.108 & -4.270 & 1.00E+05    & A & Si I \\  
 15568.325 & 5.883 & -2.447 & 1.14E+08    & A & Fe I \\  
 15568.383 & 4.667 & -3.512 & 1.49E+08    & A & V  I \\  
 15569.103 & 7.113 &-3.470 & 1.00E+05 & 1.30 & Si I \\  
 15569.240 & 5.512 & -2.360 & 1.48E+08    & A & Fe I \\  
 15569.741 & 5.750 & -1.122 & 1.43E+08    & A & Co I \\  
 15570.202 & 8.416 & -2.469 & 4.69E+08    & A & Ni II  \\
 15571.099 & 4.681 & -0.285 & 1.62E+08    & A & V  I \\  
 15571.120 & 5.879 & -1.690 & 1.75E+08    & A & Fe I \\  
 15571.729 & 2.575 & -6.048 & 1.61E+06    & A & V  I \\  
 15571.740 & 6.320 & -0.900 & 1.00E+05    & A & Fe I \\  
 15572.166 & 5.541 & -3.128 & 2.00E+08    & A & Sc I \\  
 15572.632 & 4.669 & -0.028 & 1.87E+08    & A & Ti I \\  
 15572.651 & 4.649 & -2.634 & 1.24E+08    & A & V  I \\  
 15573.976 & 5.936 & -2.790 & 7.24E+07 & 2.50 & Cr I \\  
 15574.060 & 6.310 & -1.440 & 1.00E+05    & A & Fe I \\  
 15576.040 & 5.507 & -2.270 & 2.49E+08    & A & Fe I \\  
 15576.528 & 5.169 & -3.362 & 3.67E+08    & A & Ti I \\  
 15577.594 & 5.250 & -6.311 & 1.18E+08    & A & Ca I \\  
 15577.861 & 5.250 & -3.983 & 1.18E+08    & A & Ca I \\  
 15579.080 & 6.320 & -1.05  & 1.00E+05    & A & Fe I & * \\
 15579.568 & 4.726 & -1.553 & 2.52E+08    & A & V  I \\  
 15580.150 & 6.149 & -1.887 & 1.15E+08    & A & Mn I \\  
 15580.434 & 5.250 & -1.210 & 1.54E+08    & A & Ca I \\  
 15580.876 & 7.125 & -2.910 & 1.00E+05    & A & Si I \\  
 15582.333 & 4.409 & -1.376 & 1.11E+08    & A & Ti I \\  
 15582.987 & 5.796 & -4.133 & 2.07E+08    & A & Fe I \\  
 15583.793 & 4.968 & -1.804 & 9.75E+07    & A & Sc I \\  
 15584.619 & 5.021 & -2.579 & 5.53E+07 & 2.50 & Sc I \\  
 15585.360 & 6.360 & -3.139 & 1.83E+08    & A & Fe I \\  
 15585.528 & 6.285 & -2.755 & 2.05E+08    & A & Ni I \\  
 15585.785 & 8.098 & -1.623 & 4.71E+08    & A & Ti II  \\
 15587.195 & 5.766 & -3.053 & 2.67E+08    & A & Co I \\  
 15587.282 & 4.396 & -1.528 & 1.23E+08    & A & Ti I \\  
 15587.316 & 4.958 & -1.608 & 6.85E+07    & A & Sc I \\  
 15588.260 & 5.490 & -2.70  & 1.00E+05    & A & Fe I & * \\
 15588.260 & 6.370 &  0.34  & 1.00E+05    & A & Fe I & * \\
 15588.626 & 5.491 & -2.759 & 1.69E+08    & A & Fe I \\  
 15588.699 & 6.484 & -3.638 & 3.58E+08    & A & Fe I \\  
 15588.947 & 5.447 & -2.849 & 1.19E+08    & A & Ni I \\  
 15589.139 & 4.865 & -2.506 & 1.21E+08 & 2.50 & V  I \\  
 15589.455 & 5.140 & -3.774 & 1.17E+08    & A & Ti I \\  
 15589.626 & 5.175 & -4.426 & 3.72E+08    & A & Co II  \\
 15589.893 & 4.986 & -0.988 & 7.60E+07 & 2.50 & Sc I \\  
 15590.050 & 6.240 & -0.430 & 1.00E+05    & A & Fe I \\  
 15590.301 & 4.106 & -2.583 & 1.24E+08    & A & Ti I \\  
 15590.914 & 4.191 & -3.859 & 1.78E+07    & A & Sc I \\  
 15591.490 & 6.240 &  0.450 & 1.00E+05    & A & Fe I \\  
 15591.490 & 6.360 &  0.450 & 1.00E+05    & A & Fe I \\  
 15591.595 & 6.242 &  0.129 & 1.93E+08    & A & Fe I \\  
 15591.717 & 6.045 & -1.924 & 1.46E+08    & A & Co I \\  
 15591.841 & 6.269 & -3.640 & 1.00E+05    & A & Si I \\  
 15592.762 & 4.664 & -0.838 & 1.07E+08 & 2.50 & V  I \\  
 15592.859 & 4.954 & -1.832 & 1.40E+08    & A & Sc I \\  
 15592.874 & 4.411 & -2.420 & 1.69E+08    & A & Ti I \\  
 15593.149 & 4.382 & -3.184 & 1.20E+07    & A & Ti I \\  
 15593.202 & 5.879 & -3.570 & 1.13E+08    & A & Fe I \\  
 15593.718 & 5.026 & -1.353 & 2.03E+07    & A & Ca I \\  
 15593.740 & 5.033 & -1.92  & 5.96E+08    & A & Fe I & * \\
 15594.076 & 4.965 & -2.999 & 3.46E+07 & 2.50 & Sc I \\  
 15594.903 & 2.871 & -6.423 & 1.06E+07    & A & Co I \\  
 15595.878 & 4.780 & -4.715 & 1.58E+08    & A & Ca I \\  
 15596.114 & 5.796 & -2.843 & 2.09E+08    & A & Fe I \\  
 15596.192 & 4.090 & -1.413 & 3.13E+08    & A & V  I \\  
 15596.557 & 4.965 & -0.725 & 6.82E+07    & A & Sc I \\  
 15597.163 & 5.410 & -3.913 & 1.69E+08    & A & Fe I \\  
 15597.326 & 6.726 & -2.180 & 1.00E+05    & A & Mg I \\  
 15598.870 & 6.240 & -0.880 & 1.00E+05    & A & Fe I \\  
 15598.890 & 4.690 & -0.030 & 1.00E+05    & A & Ti I \\  
 15599.134 & 4.690 & -0.250 & 1.04E+08    & A & Ti I \\  
 15599.210 & 5.948 & -2.043 & 1.54E+08 & 2.50 & Cr I \\  
 15599.346 & 4.475 & -1.948 & 7.23E+07    & A & Ti I \\  
 15600.208 & 4.684 & -1.577 & 1.36E+08    & A & Sc I \\  
 15600.500 & 5.011 & -9.268 & 2.49E+08    & A & Cr II  \\
 15600.972 & 5.228 & -6.704 & 6.43E+07    & A & Ca I \\  
 15601.810 & 5.229 & -6.258 & 6.41E+07    & A & Ca I \\  
 15601.839 & 5.773 & -3.538 & 1.47E+08    & A & Co I \\  
 15602.253 & 4.865 & -1.550 & 1.25E+08    & A & V  I \\  
 15602.496 & 5.239 & -3.253 & 5.14E+07    & A & Cr I \\  
 15602.840 & 2.267 & -1.70  & 2.31E+06    & A & Ti I & * \\
 15603.753 & 6.727 & -2.660 & 1.00E+05    & A & Mg I \\  
 15603.882 & 5.303 & -2.700 & 1.33E+08    & A & Ni I \\  
 15604.220 & 6.240 &  0.49  & 1.00E+05    & A & Fe I & * \\
 15605.073 & 2.345 & -3.464 & 7.13E+07    & A & Ti I \\  
 15605.680 & 5.300 & -0.52  & 1.00E+05    & A & Ni I & * \\
 15605.680 & 5.300 & -1.010 & 1.00E+05    & A & Ni I \\  
 15605.865 & 4.780 & -5.161 & 1.58E+08    & A & Ca I \\  
 15606.004 & 4.534 & -1.544 & 3.45E+07    & A & Sc I \\  
 15606.077 & 3.753 & -3.902 & 2.50E+08    & A & V  II  \\
 15606.661 & 8.599 & -2.301 & 5.86E+08    & A & V  II  \\
 15607.587 & 6.149 & -2.192 & 1.15E+08    & A & Mn I \\  
 15607.792 & 4.495 & -8.994 & 3.05E+08    & A & Fe II  \\
 15607.802 & 4.877 & -3.048 & 8.93E+07    & A & Ca I \\  
 15608.001 & 5.391 & -8.875 & 2.71E+08    & A & Mn II  \\
 15609.366 & 4.953 & -0.336 & 7.00E+07    & A & Sc I \\  
 15609.995 & 5.168 & -3.409 & 9.62E+07    & A & Ca I \\  
 15611.150 & 3.415 & -3.12  & 1.57E+07    & A & Fe I & * \\
 15611.219 & 6.019 & -2.362 & 3.27E+08    & A & Co I \\  
 15612.530 & 4.877 & -2.568 & 8.91E+07    & A & Ca I \\  
 15612.755 & 4.726 & -4.292 & 5.89E+07    & A & Mn I \\  
 15613.554 & 5.931 & -2.090 & 1.05E+08    & A & Fe I \\  
 15613.630 & 6.350 & -0.210 & 1.00E+05    & A & Fe I \\  
 15613.657 & 2.122 & -3.503 & 5.66E+06    & A & V  I \\  
 15614.100 & 6.350 & -0.42  & 1.00E+05    & A & Fe I & * \\
 15614.345 & 5.931 & -4.033 & 1.07E+08    & A & Fe I \\  
 15614.442 & 4.798 & -2.107 & 2.16E+08    & A & Ti I \\  
 15615.047 & 2.505 & -4.437 & 1.24E+08    & A & V  I \\  
 15615.071 & 4.057 & -3.953 & 6.75E+07    & A & V  I \\  
 15617.601 & 7.113 & -2.410 & 1.00E+05 & 1.30 & Si I \\  
 15617.828 & 5.812 & -4.036 & 2.64E+08    & A & Cr I \\  
 15618.308 & 7.113 & -2.630 & 1.00E+05 & 1.30 & Si I \\  
 15619.265 & 5.458 & -3.766 & 1.06E+08    & A & Fe I \\  
 15619.958 & 2.294 & -0.600 & 1.00E+05    & A & Y  I \\  
 15620.000 & 5.955 & -2.692 & 3.06E+08    & A & Co I \\  
 15620.610 & 5.002 & -3.468 & 3.58E+08    & A & Co I \\  
 15621.059 & 4.420 & -1.755 & 2.18E+08    & A & Ti I \\  
 15621.230 & 6.366 & -3.375 & 2.42E+08    & A & Ni I \\  
 15621.245 & 8.537 & -4.000 & 1.00E+05    & A & C  I \\  
 15621.664 & 5.539 &  0.42  & 1.12E+08    & A & Fe I & * \\
 15622.294 & 3.799 & -2.606 & 2.45E+08    & A & V  II  \\
 15622.416 & 8.082 & -0.595 & 5.19E+08    & A & Ti II  \\
 15622.450 & 5.228 & -5.071 & 5.81E+07    & A & Ca I \\  
 15623.290 & 5.229 & -3.535 & 5.79E+07    & A & Ca I \\  
 15623.319 & 8.537 & -3.160 & 1.00E+05    & A & C  I \\  
 15623.817 & 5.228 & -3.510 & 5.81E+07    & A & Ca I \\  
 15624.003 & 2.578 & -4.991 & 1.61E+06    & A & V  I \\  
 15624.140 & 4.878 & -2.342 & 8.89E+07    & A & Ca I \\  
 15624.433 & 5.229 & -2.483 & 5.79E+07    & A & Ca I \\  
 15624.526 & 5.228 & -2.765 & 5.81E+07    & A & Ca I \\  
 15624.657 & 5.229 & -2.614 & 5.81E+07    & A & Ca I \\  
 15624.726 & 6.027 & -0.428 & 8.51E+07    & A & B  I \\  
 15625.688 & 4.402 & -2.193 & 2.24E+08    & A & Ti I \\  
 15626.494 & 8.537 & -2.970 & 1.00E+05    & A & C  I \\  
 15626.799 & 4.798 & -1.400 & 2.17E+08    & A & Ti I \\  
 15628.057 & 5.791 & -3.081 & 2.16E+08    & A & Mn I \\  
 15628.180 & 5.791 & -4.060 & 1.87E+08    & A & Co I \\  
 15628.448 & 1.950 & -3.157 & 4.94E+06    & A & V  I \\  
 15629.081 & 6.027 & -0.128 & 8.51E+07    & A & B  I \\  
 15629.370 & 5.946 & -1.670 & 1.29E+08    & A & Fe I \\  
 15629.384 & 5.946 & -3.767 & 1.29E+08    & A & Fe I \\  
 15629.630 & 4.559 & -3.08  & 2.72E+08    & A & Fe I & * \\
 15630.354 & 5.791 & -3.259 & 2.15E+08    & A & Mn I \\  
 15630.916 & 4.592 & -1.275 & 1.85E+08    & A & Sc I \\  
 15631.112 & 3.642 & -4.10  & 8.04E+07    & A & Fe I & * \\
 15631.430 & 4.777 & -3.590 & 4.15E+07    & A & Ti I \\  
 15631.476 & 7.125 & -2.140 & 1.00E+05    & A & Si I \\  
 15631.845 & 5.033 & -2.109 & 5.58E+07 & 2.50 & Sc I \\  
 15631.935 & 8.307 & -4.217 & 4.97E+08    & A & Fe II  \\
 15631.950 & 5.352 &  0.15  & 1.14E+08    & A & Fe I & * \\
 15632.654 & 5.305 & -0.01  & 9.71E+07    & A & Ni I & * \\
 15633.079 & 5.045 & -6.184 & 3.88E+07    & A & Ca I \\  
 15633.996 & 4.073 & -1.587 & 3.13E+08    & A & V  I \\  
 15634.730 & 6.054 & -4.151 & 7.18E+07 & 2.50 & Co I \\  
 15635.317 & 4.942 & -2.073 & 1.36E+08    & A & Sc I \\  
 15635.708 & 5.161 & -2.323 & 1.18E+08    & A & Ti I \\  
 15636.292 & 4.616 & -8.153 & 3.08E+08    & A & Fe II  \\
 15636.803 & 4.798 & -2.010 & 2.17E+08    & A & Ti I \\  
 15637.082 & 5.228 & -7.521 & 1.87E+08    & A & Ca I \\  
 15637.214 & 6.724 & -2.763 & 4.50E+08    & A & Fe II  \\
 15637.924 & 5.229 & -6.571 & 1.87E+08    & A & Ca I \\  
 15637.965 & 6.361 & -2.20  & 2.33E+08    & A & Fe I & * \\
 15638.029 & 5.507 & -2.797 & 2.34E+08    & A & Fe I \\  
 15638.300 & 6.076 & -4.458 & 2.31E+08    & A & Cr II  \\
 15638.472 & 6.734 & -1.93  & 1.00E+05    & A & Si I & * \\
 15638.919 & 5.814 & -1.74  & 1.41E+08    & A & Fe I & * \\
 15639.480 & 6.410 & -0.88  & 1.00E+05    & A & Fe I & * \\
 15640.869 & 6.212 & -1.063 & 2.61E+08    & A & Mn I \\  
 15640.872 & 6.098 & -2.933 & 1.40E+08    & A & Ni I \\  
 15641.002 & 5.539 & -3.966 & 8.93E+07    & A & Fe I \\  
 15641.701 & 1.931 & -9.488 & 4.85E+06    & A & V  I \\  
 15641.922 & 5.494 & -3.792 & 4.92E+08    & A & Ni I \\  
 15642.338 & 5.303 & -3.107 & 1.32E+08    & A & Ni I \\  
 15643.072 & 6.317 & -3.075 & 3.10E+08    & A & V  II  \\
 15643.219 & 8.548 & -2.870 & 7.45E+08    & A & V  II  \\
 15643.380 & 6.783 & -2.350 & 1.00E+05    & A & Mg I \\  
 15645.010 & 6.310 & -0.57  & 1.00E+05    & A & Fe I & * \\
 15645.378 & 4.993 & -1.720 & 8.71E+06    & A & Al I \\  
 15645.415 & 7.113 & -1.630 & 1.00E+05 & 1.30 & Si I  \\  
 15645.594 & 5.674 & -3.812 & 1.00E+08    & A & Co I \\  
 15645.843 & 7.166 & -4.450 & 1.00E+05    & A & Si I \\  
 15646.940 & 4.958 & -3.147 & 3.43E+07 & 2.50 & Sc I \\  
 15646.965 & 5.800 & -2.066 & 1.24E+08    & A & Cr I \\  
 15647.332 & 4.725 & -3.269 & 1.72E+08    & A & V  I \\  
 15647.410 & 6.330 & -1.05  & 1.00E+05    & A & Fe I & * \\
 15647.479 & 4.059 & -1.709 & 2.72E+08    & A & V  I \\  
 15648.535 & 5.426 & -0.66  & 1.12E+08    & A & Fe I & * \\
 15649.750 & 5.550 & -3.027 & 3.24E+08    & A & Fe I \\  
 15651.154 & 4.681 & -3.859 & 1.80E+08    & A & V  I \\  
 15652.354 & 5.744 & -2.363 & 3.12E+08    & A & Co I \\  
 15652.870 & 6.250 & -0.13  & 1.00E+05    & A & Fe I & * \\
 15652.930 & 2.482 & -9.038 & 4.40E+03    & A & Fe I \\  
 15653.457 & 5.464 & -4.076 & 8.04E+07 & 2.50 & V  I \\  
 15656.473 & 4.591 & -3.391 & 1.10E+08 & 2.50 & V  I \\  
 15656.659 & 5.828 & -2.847 & 1.26E+08    & A & Fe I \\  
 15656.669 & 5.874 & -1.80  & 1.41E+08    & A & Fe I & * \\
 15659.317 & 5.537 & -3.456 & 1.22E+08    & A & Co I \\  
 15659.465 & 4.389 & -2.263 & 1.83E+08    & A & Cr I \\  
 15659.729 & 6.779 & -4.050 & 1.00E+05    & A & Mg I \\  
 15659.771 & 5.538 & -3.001 & 2.97E+08    & A & Fe I \\  
 15659.855 & 4.994 & -1.420 & 8.71E+06    & A & Al I \\  
 15660.259 & 6.779 & -3.450 & 1.00E+05    & A & Mg I \\  
 15660.375 & 6.779 & -2.060 & 1.00E+05    & A & Mg I \\  
 15661.290 & 5.607 & -2.631 & 1.50E+08    & A & Fe I \\  
 15661.555 & 4.506 & -3.327 & 2.04E+08    & A & Ti I \\  
 15661.611 & 4.795 & -3.961 & 2.06E+08    & A & Fe I \\  
 15661.655 & 4.650 &  0.266 & 8.83E+07    & A & V  I \\  
 15662.010 & 5.830 &  0.25  & 1.00E+05    & A & Fe I & * \\
 15662.320 & 6.330 & -0.80  & 1.00E+05    & A & Fe I & * \\
 15662.686 & 4.850 & -2.301 & 1.83E+08    & A & Ti I \\  
 15662.826 & 3.087 & -6.216 & 7.82E+06    & A & Cr I \\  
 15663.047 & 5.140 & -3.605 & 1.19E+08    & A & Ti I \\  
 15663.905 & 5.102 & -1.972 & 6.41E+07 & 2.50 & Sc I \\  
 15664.735 & 4.370 & -1.742 & 8.11E+07    & A & Ti I \\  
 15665.240 & 5.980 & -0.600 & 1.00E+05    & A & Fe I \\  
 15665.994 & 5.233 & -2.344 & 2.11E+08    & A & Ti I \\  
 15666.299 & 4.411 & -2.121 & 2.27E+08    & A & Ti I \\  
 15666.679 & 5.520 & -3.484 & 7.18E+07    & A & Mn I \\  
 15666.728 & 3.096 & -5.656 & 2.84E+03    & A & Cr I \\  
 15667.956 & 4.888 & -3.803 & 1.07E+08    & A & Sc I \\  
 15670.130 & 6.200 & -1.02  & 1.00E+05    & A & Fe I & * \\
 15670.387 & 5.208 & -5.097 & 3.73E+08    & A & Co II  \\
 15670.977 & 4.953 & -1.130 & 6.97E+07    & A & Sc I \\  
 15671.000 & 6.330 & -0.57  & 1.00E+05    & A & Fe I & * \\
 15671.714 & 3.807 & -1.851 & 3.16E+08    & A & Sc I \\  
 15671.860 & 5.920 & -1.40  & 1.00E+05    & A & Fe I & * \\
 15672.156 & 5.611 & -4.110 & 1.67E+08    & A & Mn I \\  
 15672.574 & 5.941 & -2.701 & 9.23E+07 & 2.50 & Cr I \\  
 15673.150 & 6.250 & -0.73  & 1.00E+05    & A & Fe I & * \\
 15673.385 & 5.133 & -0.57  & 6.68E+07    & A & Mn I & * \\
 15673.982 & 5.883 & -2.826 & 1.12E+08    & A & Fe I \\  
 15674.122 & 6.022 & -2.537 & 4.30E+08    & A & Co I \\  
 15674.392 & 5.382 & -3.351 & 1.45E+08    & A & V  I \\  
 15674.653 & 7.064 & -1.30  & 1.00E+05    & A & Si I & * \\
 15676.277 & 7.092 & -4.110 & 1.00E+05    & A & Si I \\  
 15676.599 & 5.106 & -1.85  & 8.13E+07    & A & Fe I & * \\
 15676.825 & 5.521 & -2.786 & 1.25E+08 & 2.50 & V  I \\  
 15677.020 & 6.250 & -0.730 & 1.00E+05    & A & Fe I \\  
 15677.520 & 6.250 &  0.20  & 1.00E+05    & A & Fe I & * \\
 15677.588 & 5.400 & -2.641 & 1.26E+08    & A & V  I \\  
 15678.340 & 5.829 & -2.030 & 2.16E+08    & A & Fe I \\  
 15678.519 & 7.125 & -2.40  & 1.00E+05    & A & Si I & * \\
 15678.989 & 9.045 & -3.905 & 8.30E+08    & A & V  II  \\
 15679.373 & 5.872 & -3.699 & 1.15E+08    & A & Fe I \\  
 15679.983 & 6.274 & -3.160 & 1.00E+05    & A & Si I \\  
 15680.081 & 4.697 &  0.10  & 2.04E+08    & A & Cr I & * \\
 15680.194 & 5.094 & -1.851 & 1.54E+08    & A & Ti I \\  
 15680.863 & 5.297 & -2.705 & 1.52E+08    & A & Ti I \\  
 15681.658 & 2.851 & -9.416 & 4.97E+04    & A & Fe I \\  
 15682.020 & 5.410 & -2.230 & 1.12E+08    & A & Fe I \\  
 15682.510 & 6.370 & -0.40  & 1.00E+05    & A & Fe I & * \\
 15682.826 & 2.581 & -4.766 & 1.61E+06    & A & V  I \\  
 15683.390 & 5.621 & -1.970 & 1.47E+08    & A & Fe I \\  
 15683.490 & 4.741 & -2.909 & 1.46E+08    & A & V  I \\  
 15685.168 & 2.318 & -5.865 & 1.72E+06    & A & Ti I \\  
 15685.375 & 4.405 & -2.849 & 2.24E+08    & A & Ti I \\  
 15685.407 & 5.326 & -3.206 & 1.27E+08    & A & Ni I \\  
 15685.604 & 5.913 & -2.700 & 1.20E+08    & A & Fe I \\  
 15686.020 & 6.330 & -0.20  & 1.00E+05    & A & Fe I & * \\
 15686.165 & 5.913 & -3.760 & 1.23E+08    & A & Fe I \\  
 15686.222 & 4.664 &  0.430 & 8.99E+07 & 2.50 & V  I \\  
 15686.271 & 5.104 & -3.404 & 6.43E+07 & 2.50 & Sc I \\  
 15686.440 & 6.250 &  0.17  & 1.00E+05    & A & Fe I & * \\
 15686.449 & 5.699 & -3.009 & 3.63E+07    & A & Cr I \\  
 15686.449 & 5.699 & -3.118 & 3.52E+07    & A & Cr I \\  
 15688.265 & 5.839 & -4.013 & 1.20E+08 & 2.50 & Co I \\  
 15689.225 & 4.650 & -3.940 & 2.49E+08    & A & V  I \\  
 15691.850 & 6.250 &  0.61  & 1.00E+05    & A & Fe I & * \\
 15692.750 & 5.385 & -0.50  & 1.14E+08    & A & Fe I & * \\
 15692.781 & 3.397 & -5.402 & 1.50E+07    & A & Fe I \\  
 15692.845 & 5.133 & -0.827 & 6.89E+07    & A & Mn I \\  
 15692.993 & 7.482 & -0.980 & 4.47E+08    & A & Sc II  \\
 15693.166 & 4.650 & -4.064 & 2.42E+08    & A & V  I \\  
 15693.360 & 6.719 & -1.20  & 1.00E+05    & A & Mg I & * \\
 15693.360 & 6.719 & -2.070 & 1.00E+05    & A & Mg I \\  
 15693.360 & 6.719 & -3.620 & 1.00E+05    & A & Mg I \\  
 15693.454 & 6.719 & -1.180 & 1.00E+05    & A & Mg I \\  
 15693.454 & 6.719 & -2.080 & 1.00E+05    & A & Mg I \\  
 15693.555 & 6.719 & -1.340 & 1.00E+05    & A & Mg I \\  
 15693.782 & 5.497 & -3.360 & 3.32E+08    & A & Ni I \\  
 15693.929 & 5.161 & -3.028 & 1.19E+08    & A & Ti I \\  
 15694.500 & 6.240 & -1.520 & 1.00E+05    & A & Fe I \\  
 15694.964 & 5.219 & -2.606 & 3.37E+08    & A & Ti I \\  
 15696.319 & 2.616 & -2.439 & 1.68E+06    & A & V  I \\  
 15696.371 & 4.396 & -2.038 & 1.92E+08    & A & Ti I \\  
 15696.472 & 5.156 & -1.164 & 2.54E+08    & A & Ti I \\  
 15696.911 & 5.538 & -2.280 & 1.19E+08    & A & Cr I \\  
 15698.508 & 5.519 & -2.393 & 1.52E+08    & A & Fe I \\  
 15698.765 & 5.476 & -2.405 & 1.43E+08    & A & Fe I \\  
 15698.979 & 1.887 & -2.09  & 2.76E+06    & A & Ti I & * \\
 15699.129 & 3.803 & -3.855 & 2.45E+08    & A & V  II  \\
 15700.090 & 6.330 & -1.180 & 1.00E+05    & A & Fe I \\  
 15700.685 & 5.228 & -7.463 & 1.86E+08    & A & Ca I \\  
 15700.806 & 5.560 & -3.905 & 9.62E+07 & 2.50 & V  I \\  
 15701.341 & 5.049 & -7.810 & 7.66E+07    & A & Ca I \\  
 15701.447 & 3.758 & -2.916 & 2.50E+08    & A & V  II  \\
 15702.088 & 7.855 & -3.000 & 3.59E+08    & A & V  II  \\
 15702.882 & 5.203 & -2.348 & 2.64E+08    & A & Ti I \\  
 15704.111 & 4.534 & -2.852 & 3.28E+07    & A & Sc I \\  
 15704.432 & 3.369 & -2.618 & 4.05E+06    & A & Cr I \\  
 15704.654 & 8.116 & -1.903 & 4.76E+08    & A & Ti II  \\
 15704.802 & 5.991 & -3.081 & 3.40E+08    & A & Co I \\  
 15705.312 & 5.049 & -5.896 & 7.66E+07    & A & Ca I \\  
 15706.183 & 5.330 & -7.863 & 2.35E+08    & A & Cr II  \\
 15707.047 & 4.958 & -0.437 & 6.89E+07    & A & Sc I \\  
 15708.380 & 4.196 & -3.193 & 1.77E+07    & A & Sc I \\  
 \end{longtable}


\end{document}